\documentclass[twocolumn]{aastex62}

\usepackage{graphicx}
\usepackage{amsmath}	
\usepackage{amssymb}
\usepackage{booktabs}
\usepackage{enumerate}
\usepackage{dirtytalk}
\usepackage{float}
\usepackage{enumitem}
\usepackage{booktabs}
\usepackage{lipsum} 
\usepackage{float}
\usepackage{xcolor}
\usepackage{multirow}
\usepackage{makecell}
\usepackage[english]{babel}

\received{XX XX XXXX, ApJ}

\shorttitle{\textsc{Aura-3D}: Three-dimensional Atmospheric Retrieval}
\shortauthors{Nixon \& Madhusudhan}

\begin{document}

\title{\textsc{Aura-3D}: A Three-dimensional Atmospheric Retrieval Framework for Exoplanet Transmission Spectra}

\author{Matthew C. Nixon}
\email{mnixon@ast.cam.ac.uk}
\affil{Institute of Astronomy, University of Cambridge, Madingley Road, Cambridge CB3 0HA, UK}

\author{Nikku Madhusudhan}
\email{nmadhu@ast.cam.ac.uk}
\affil{Institute of Astronomy, University of Cambridge, Madingley Road, Cambridge CB3 0HA, UK}

\begin{abstract}
Atmospheric retrievals of exoplanet transmission spectra allow constraints on the composition and structure of the day-night terminator region. Such retrievals in the past have typically assumed one-dimensional temperature structures which were adequate to explain extant observations. However, the increasing data quality expected from exoplanet spectroscopy with JWST motivates considerations of multidimensional atmospheric retrievals. We present \textsc{Aura-3D}, a three-dimensional atmospheric retrieval framework for exoplanet transmission spectra. \textsc{Aura-3D} includes a forward model that enables rapid computation of transmission spectra in 3D geometry for a given atmospheric structure and can, therefore, be used for atmospheric retrievals as well as for computing spectra from General Circulation Models (GCMs). In order to efficiently explore the space of possible 3D temperature structures in retrievals, we develop a parametric 3D pressure-temperature profile which can accurately represent azimuthally-averaged temperature structures of a range of hot Jupiter GCMs. We apply our retrieval framework to simulated JWST observations of hot Jupiter transmission spectra, obtaining accurate estimates of the day-night temperature variation across the terminator as well as the abundances of chemical species. We demonstrate an example of a model hot Jupiter transmission spectrum for which a traditional 1D retrieval of JWST-quality data returns biased abundance estimates, whereas a retrieval including a day-night temperature gradient can accurately retrieve the true abundances. Our forward model also has the capability to include inhomogeneous chemistry as well as variable clouds/hazes. This new retrieval framework opens the field to detailed multidimensional atmospheric characterisation using transmission spectra of exoplanets in the JWST era.
\end{abstract}




\section{Introduction}
\label{section:intro}

\begin{figure*}
\centering
\includegraphics[width=0.9\linewidth,trim={0 5cm 4.4cm 0},clip]{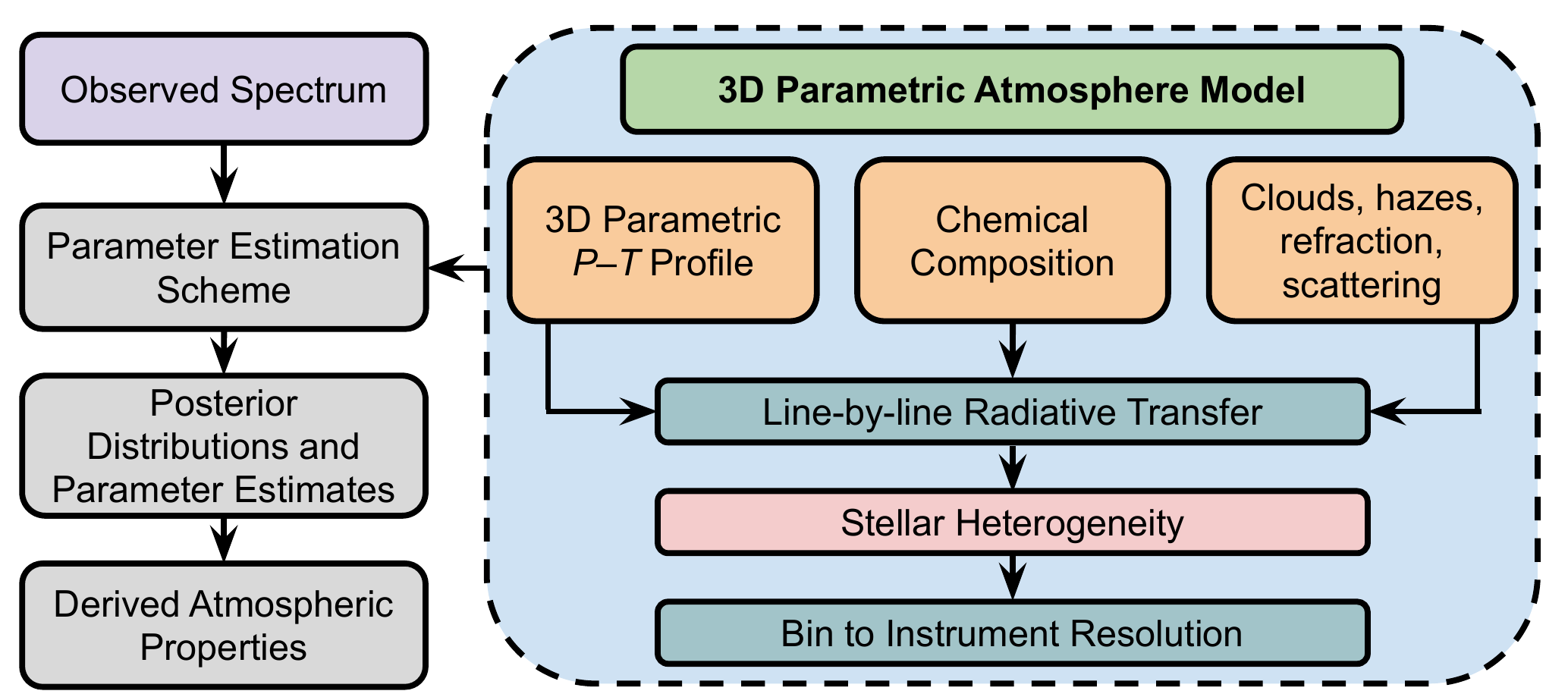}
    \caption{Schematic of the \textsc{Aura-3D} retrieval framework. The algorithm combines an atmospheric forward model in 3D geometry with a Bayesian sampling algorithm to conduct parameter estimation. The forward model is modular in structure, with numerous features that can be incorporated as needed, such as clouds/hazes, refraction, scattering and stellar heterogeneity.}
    \label{fig:retrieval_schematic}
\end{figure*}

The study of exoplanet atmospheres has seen immense progress in recent years, with spectroscopic observations enabling constraints on a range of physical and chemical processes that occur in exoplanets \citep{Madhu2019}. One of the most successful observational techniques used to characterise exoplanet atmospheres has been transmission spectroscopy, which measures the wavelength-dependent decrease in flux from the host star as the planet transits in front of it. This approach has led to detections and abundance constraints of chemical species in numerous exoplanets \citep[e.g.,][]{Charbonneau2002,Snellen2010,Deming2013,Mandell2013,Kreidberg2014,Madhu2014,Wyttenbach2015,Wakeford2018,Pinhas2019} as well as providing constraints on the properties of clouds and hazes \citep[e.g.,][]{Pont2008,Bean2010,Pont2013,Kreidberg2014_GJ,Sing2016,Barstow2017,Pinhas2019,Benneke2019}.

One of the most common methods for deriving atmospheric properties from spectroscopic data of exoplanets is atmospheric retrieval \citep{Madhu2009}. An atmospheric retrieval algorithm involves combining an atmospheric forward model with a statistical sampling algorithm in order to determine estimates of the atmospheric properties that best explain the observed data \citep[see e.g.,][for a recent review]{Madhu2018}. The earliest retrieval algorithms employed a grid-based search to find best-fitting models \citep{Madhu2009}, with later studies incorporating Bayesian statistical inference schemes such as Markov Chain Monte Carlo \citep[e.g.,][]{Madhu2010,Line2013,Cubillos2013,Zhang2019,Lacy2020}, and Nested Sampling \citep[e.g.,][]{Benneke2013,Waldmann2015,Oreshenko2017,MacDonald2017,Gandhi2018,Molliere2019,Zhang2020}. Other approaches to parameter estimation, such as Optimal Estimation \citep[e.g.,][]{Irwin2008,Lee2012,Barstow2017} and Machine Learning \citep[e.g.,][]{MarquezNeila2018,Zingales2018,Cobb2019,Fisher2020,Nixon2020}, have also been implemented.

In order to create an accurate and flexible atmospheric model for use in retrievals, a large number of parameters are required to capture all possible compositions, thermal structures, and other properties such as the presence of clouds or hazes. Since statistical sampling algorithms often require a large number of model evaluations, increasing as the number of free parameters increases \citep{Fortney2021}, it is necessary to make some simplifying assumptions to reduce computation time. Common assumptions include the use of a 1D temperature profile and uniform chemical abundances in the region of the atmosphere probed by the observations \citep{Madhu2018}. It is important to re-examine the assumptions made in these algorithms as we enter a new era in the study of exoplanet atmospheres, in which the quality and volume of observational data will improve immensely thanks to next-generation facilities such as the \textit{James Webb} Space Telescope \citep[JWST;][]{Greene2016}, the \textit{Atmospheric Remote-sensing Infrared Exoplanet Large-survey} \citep[ARIEL;][]{Tinetti2018} and the Extremely Large Telescopes \citep[ELTs;][]{Snellen2015}.

Despite the successes of the retrieval approach for analysing data from existing observatories such as the Hubble Space Telescope (HST), several studies have highlighted the potential for 1D retrieval codes to lead to biased results due to variations in temperature structure and chemical abundances with latitude and longitude, particularly when considering high-quality observations of very hot Jupiters from upcoming facilities such as JWST. For example, assumptions of a 1D atmosphere have been shown to bias retrievals of emission spectra \citep[e.g.,][]{Blecic2017,Taylor2020} and can lead to unrealistic model fits to phase curves \citep[e.g.,][]{Irwin2020,Feng2020}.

This work focuses on retrievals of transmission spectra. Transmission spectroscopy probes the day-night terminator region of the atmosphere, across which inhomogeneities may be expected \citep[e.g.,][]{Fortney2010}. These effects should be particularly prevalent in hot and ultra-hot Jupiters, since day-night temperature contrasts are expected to increase as planetary equilibrium temperature increases \citep{Cowan2011,Komacek2016}. Previous studies have already indicated that 1D retrievals can lead to biases when interpreting transmission spectra of hot giant planets. \citet{Caldas2019} developed a transmission spectrum model that could incorporate a 3D atmospheric structure, which they used to investigate the effect of day-night temperature gradients on model spectra. They used the 1D retrieval code \textsc{TauREx} \citep{Waldmann2015} to analyse synthetic JWST spectra generated with their 3D model for which the temperature could vary significantly between the day- and nightsides. The retrievals returned terminator temperatures that were biased towards the dayside temperature, and chemical abundances which were not consistent with input values. \citet{Pluriel2022} also found day-night induced biases for a range of synthetic JWST-like spectra of hot and ultra-hot Jupiters, and \citet{Pluriel2020} used the same 3D model to explore the potential for biases caused by day–night chemical heterogeneities. They focused on the ultra-hot Jupiter WASP-121b, for which the H$_2$O abundance is expected to be much higher on the nightside than the dayside due to thermal dissociation \citep{Parmentier2018}. A 1D retrieval analysis showed that inferred chemical abundances could be significantly different to the input values. \citet{MacDonald2020} also discussed the potential for biased temperatures and abundances from 1D retrievals \citep[but cf.][]{Welbanks2022}. A generalised retrieval framework should be able to address the above issues in order to accurately infer atmospheric properties regardless of the presence of thermal inhomogeneities. This is the goal of the present study.

Unlike the simplified atmospheric models used in retrieval algorithms, General Circulation Models (GCMs) can be used to capture some of the more complex processes at work in planetary atmospheres. GCMs are detailed 3D models that simulate atmospheric dynamics, some also incorporating chemical processes along with radiative transport \citep{Showman2020}. Originally developed to simulate the Earth's atmosphere, a wide range of GCMs have been adapted for application to exoplanet atmospheres \citep{Showman2002,Cooper2005,Lewis2010,Rauscher2010,Thrastarson2010,Wordsworth2011,Polichtchouk2012,DobbsDixon2013,Kataria2013,Amundsen2016,Mendonca2016,Way2017,Deitrick2020}. These models take various different approaches to computing atmospheric dynamics. In general, GCMs aim to solve the primitive equations of atmospheric dynamics in 3D across the entire planet, which are simplifications of the Navier-Stokes equations that assume hydrostatic equilibrium, a shallow atmosphere compared to the planetary radius, and constant gravity with height. However, some models instead opt to solve the full Navier-Stokes equations \citep[e.g.,][]{DobbsDixon2013,Mayne2014,Amundsen2016}.

GCMs have predicted a number of important features of hot Jupiter atmospheres. This includes the finding that hot Jupiters can show large day-night temperature contrasts of up to serveral hundred K \citep[e.g.,][]{Showman2002,Cooper2005,Amundsen2016}. GCMs of hot Jupiters have also predicted equatorial superrotation that can lead to an eastward shift of a planet's dayside hotspot away from the substellar point \citep[e.g.,][]{Showman2002,Cooper2005,Rauscher2010,Kataria2013}. These predictions have subsequently been confirmed through comparison with observations of infrared phase curves \citep[e.g.,][]{Knutson2007,Komacek2017,Stevenson2017}.

Model transmission spectra can be generated using 3D temperature structures. \citet{Fortney2010} used pre-computed temperature profiles from SPARC/MITgcm simulations of hot Jupiters \citep{Showman2009} to examine the differences between 1D and 3D model atmospheres. A number of more recent works have also developed multidimensional forward models for transmission spectra. \citet{Caldas2019} used a 3D radiative transfer model to investigate biases caused by day-night temperature gradients using a simple parametric temperature profile. \citet{Falco2022} presented transmission spectrum models in 1D, 2D and 3D, and \citet{MacDonald2021} described parametric prescriptions for 3D atmospheric models.

Although GCMs are too computationally expensive to incorporate into retrieval algorithms directly, a number of studies have aimed to bridge the gap between 1D and 3D models in the context of retrievals of transmission spectra. \citet{Lacy2020} conducted chemical equilibrium retrievals assuming separate dayside and nightside temperature profiles, showing that in some cases it is possible to constrain day- and nightside temperatures of hot Jupiters from their transmission spectra. \citet{Espinoza2021} demonstrated how JWST could be used to acquire separate transmission spectra for each of a planet's limbs, and also expanded the retrieval framework CHIMERA \citep{Line2013} to enable retrievals of a single transmission spectrum with separate temperature profiles for each limb, under the assumption of chemical equilibrium. \citet{Welbanks2022} also presented a 1+1D retrieval framework in which the final transmission spectrum is calculated by a linear combination of two separate spectra, one representing the morning terminator and another representing the evening terminator.

In this work we present \textsc{Aura-3D}, a 3D atmospheric retrieval framework for transmission spectra of exoplanets. In Section \ref{section:methods}, we describe our algorithm in detail. We present a forward model that incorporates a 3D temperature structure and which can be used to efficiently generate transmission spectra from the output of a GCM, similarly to several past studies \citep[][]{Fortney2010,Caldas2019}. We develop a parametric pressure-temperature ($P$--$T$) profile that can match the azimuthally-averaged structure of a GCM while being very fast to compute, making it suitable for atmospheric retrieval. In Section \ref{subsec:pt_comparison} we compare our parametric temperature profile to a number of established GCMs. We explore the effects of a day-night temperature gradient on resulting transmission spectra in Section \ref{subsec:mgrid} by quantifying the difference in transit depths between 3D models and their 1D-averaged counterparts. In Section \ref{subsec:chem} we discuss the possible effects of chemical inhomogeneity on transmission spectra. In Section \ref{subsec:parametric_ret} we demonstrate the capability of \textsc{Aura-3D} to carry out retrievals with a multidimensional $P$--$T$ profile by conducting a retrieval on a synthetic hot Jupiter spectrum generated using our new parametric profile. In Section \ref{subsec:gcm_ret} we retrieve a spectrum generated using a GCM temperature structure, comparing results from retrievals both with and without a day-night temperature gradient. We summarise our findings and discuss avenues for the future in Section \ref{section:discussion}.


\section{Methods} \label{section:methods}

The retrieval framework developed in this study is shown in Figure \ref{fig:retrieval_schematic}. Similarly to previous retrieval algorithms, this framework combines an atmospheric forward model with a Bayesian parameter estimation scheme. The key differences between our methods and those of previous retrieval algorithms concern the forward model. Our forward model enables the computation of a transmission spectrum whose temperature structure can vary in three dimensions throughout the atmosphere, and can therefore take the temperature structure calculated by a GCM as its input. However, since the calculation of a temperature structure using a GCM is very computationally expensive, we also describe a number of parameterisations that allow for rapid computation of a multidimensional temperature profile, and which can be incorporated directly into a retrieval. Our retrieval framework also incorporates several components from the existing retrieval code \textsc{Aura} and its subsequent implementations \citep{WelbanksMadhu2019,Nixon2020} as well as \textit{Aurora} \citep{Welbanks2021}. These features are discussed later in this section. We note that \textit{Aurora} can also be applied to H-poor atmospheres, which are not considered in the present work.

\subsection{Transit Geometry}

\begin{figure*}
\centering
\includegraphics[width=0.8\linewidth,trim={1.8cm 3.2cm 1.2cm 0},clip]{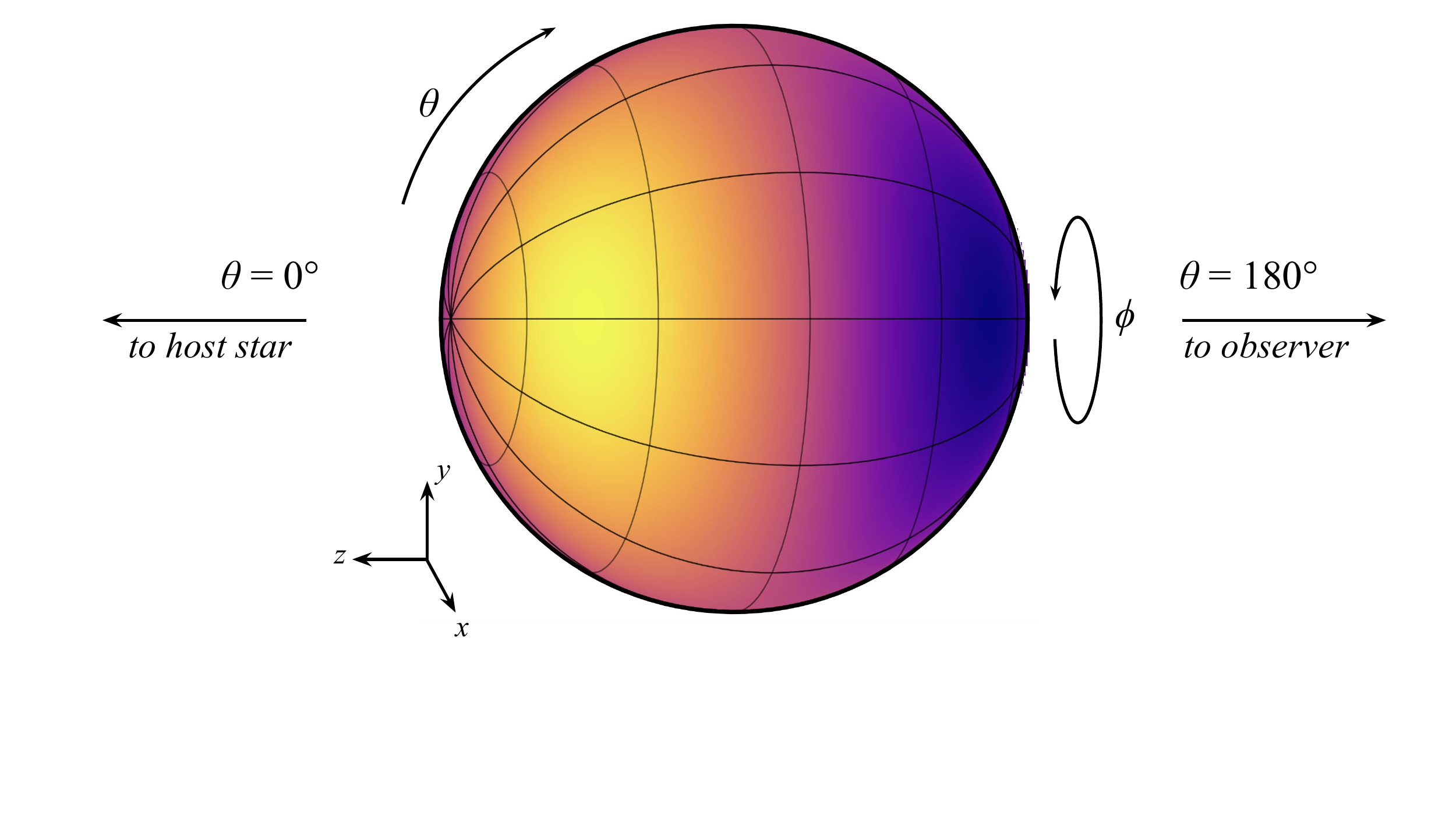}
    \caption{Three-dimensional co-ordinate system adopted for our forward model. The zenith angle $\theta$ is defined to be 0 at the substellar point and $\pi$ (180$^{\circ}$) at the antistellar point. The azimuthal angle $\phi$ varies between 0 and $2\pi$ (360$^{\circ}$), increasing in the direction of the trailing limb.}
    \label{fig:3dgeometry}
\end{figure*}

We define a spherical polar coordinate system $(r,\theta,\phi)$ with an origin at the centre of the planet (see Figure \ref{fig:3dgeometry}). The coordinates $(R_p,0,0)$ correspond to the substellar point. The zenith angle $\theta$ varies between 0 and $\pi$, reaching $\pi/2$ at the day-night terminator and $\pi$ and the antistellar point. The azimuthal angle $\phi$ varies between 0 and $2\pi$, increasing in the direction of the trailing hemisphere of the planet. This geometry, based on the star-planet-observer axis, differs from a traditional latitude-longitude grid based on the rotation axis which is often used when describing the geometry of a planet. We adopt this approach since it allows for a straightforward description of day-night variations in atmospheric properties, since the zenith angle $\theta$ moves from the substellar to antistellar point. This geometry has been used in previous studies involving transmission spectroscopy \citep[e.g.,][]{Fortney2010,Caldas2019}.

\begin{figure*}
\centering
\includegraphics[width=0.8\linewidth,trim={0.2cm 9.3cm 8cm 0cm},clip]{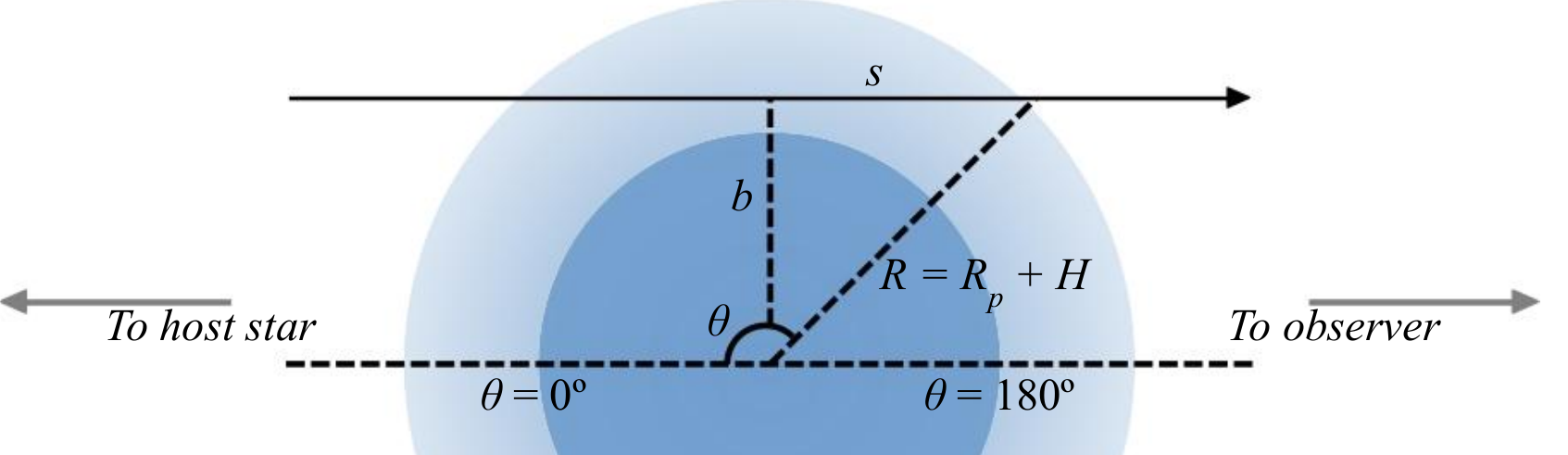}
    \caption{Geometry of an exoplanet atmosphere as observed in transit. A ray passes through the atmosphere along the path $s$, travelling through the day-night terminator region which covers a range of values of the zenith angle $\theta$.}
    \label{fig:geometry}
\end{figure*}

Figure \ref{fig:geometry} depicts a ray of light travelling from the host star to an observer, passing through a portion of the atmosphere of a transiting planet. A single ray is assumed to have a fixed impact parameter $b$ and a fixed azimuthal angle $\phi$ as it moves through the atmosphere. The ray travels along the path $s$, defined so that $s=0$ when the ray is directly above the planet's terminator ($\theta=90^{\circ}$). As it travels along this path, the zenith angle $\theta$ increases (moving from the dayside to the nightside of the planet) and the distance between the ray and the centre of the planet (denoted $r$) also varies. The values of $\theta$ and $r$ are given by

\begin{equation}
    r^2 = s^2 + b^2
\end{equation}
and
\begin{equation}
    \cos \theta = \frac{s}{r}.
\end{equation}

\subsection{Radiative Transfer}

The following expression describes the transit depth $\Delta_{\lambda}$ for the generalised atmosphere of a planet with radius $R_p$ transiting a star with radius $R_*$:

\begin{equation}
    \Delta_{\lambda} = \frac{1}{2\pi} \int_0^{2\pi} \delta_{\lambda} ({\phi}) \, \mathrm{d}\phi,
    \label{eq:td}
\end{equation}
where
\begin{align}
    \delta_{\lambda} (\phi) & = \frac{1}{R_*^2} \Bigg[ R_p^2 + 2 \int_{R_p}^{R_p+H} b \Big( b \big( 1 - e^{-\tau_{\lambda}(b,\phi)} \big)  \Big) \mathrm{d}b \nonumber \\
    & - \, 2 \int_{0}^{R_p} be^{-\tau_{\lambda}(b,\phi)} \mathrm{d}b  \Bigg].
\end{align}
In this equation, the atmospheric height is denoted by $H$, and $\tau(b,\phi)$ represents the optical depth at impact parameter $b$ and azimuthal angle $\phi$. The total transit depth $\Delta_{\lambda}$ is found by integrating the transit depth at each value of $\phi$ over all azimuthal angles. 

A crucial difference between the 3D transmission spectrum model and a 1D approximation is that the attenuation coefficient at a given wavelength, $\mu_{\lambda}$, depends on $r, \theta$ and $\phi$, as opposed to only depending on $r$. The expression for the optical depth along a ray path $s$ at a given $(b,\phi)$ is therefore
\begin{align}
    \tau_{\lambda}(b,\phi) & = \int_{-s_0}^{s_0} \mu_{\lambda}(r,\theta,\phi) \, \mathrm{d}s \nonumber \\
    & = \int_{-s_0}^{s_0} \mu_{\lambda}(s,b,\phi) \, \mathrm{d}s.
    \label{eq:tau}
\end{align}

Previous retrieval codes have included forward models with $\phi$-dependence to capture properties such as inhomogeneous cloud cover \citep{Line2016,MacDonald2017,Welbanks2021}. \textsc{Aura-3D} also incorporates $\phi$-dependence to model inhomogeneous clouds (see Section \ref{subsubsec:clouds}). By including a temperature profile that depends on $\theta$ (see \ref{subsubsec:3dpt}), our retrieval algorithm is therefore capable of constraining atmospheric properties in three dimensions.

Our model incorporates absorption from a wide range of chemical species following the methods described in \citet{Gandhi2017,Gandhi2018}, using cross-sections derived from line list data from a number of different sources. A Voigt function is used to apply temperature and pressure broadening. We assume a H/He-dominated atmosphere with solar abundances of H and He \citep{Asplund2009}. Additional chemical species present in the model which are used in this work include H$_2$O \citep{Rothman2010}, CH$_4$ \citep{Yurchenko2014}, NH$_3$ \citep{Yurchenko2011}, HCN \citep{Barber2014}, CO and CO$_2$ \citep{Rothman2010}. The attenuation coefficient of a given chemical species $i$ can be expressed as
 \begin{equation}
      \mu_i (\lambda, P, T) = \rho_i \kappa_i(\lambda, P, T) = n_i \sigma_i(\lambda,P,T), 
      \label{eq:opac}
 \end{equation}
 where $\rho_i$ is the mass density of species $i$, $\kappa_i$ is the opacity of species $i$, $n_i$ is the number density of species  $i$ and $\sigma_i$ is the absorption cross-section of species $i$. The number density $n_i$ of a given species can be related to its volume mixing ratio, $X_i=n_i/n_{\rm tot}$, where $n_{\rm tot}$ is the total number density. The volume mixing ratio of each chemical species, apart from H$_2$ and He, is a free parameter in the model. The volume mixing ratios of H$_2$ and He are given by
 \begin{align}
    X_{\rm{H}_2} &= \dfrac{1 - \sum_{i,i\neq \rm{H}_2, \rm{He}} X_i }{1 + (X_{\rm{He}}/X_{\rm{H_2}})}, \\
    X_{\rm{He}} &= 0.17 X_{\rm{H}_2} ,
\end{align}
 where the value of 0.17 is derived from a solar composition \citep{Asplund2009}.
 
 In past implementations of \textsc{Aura}, cross-sections of each species are stored on a three-dimensional grid of $\lambda, P$ and $T$. In order to determine the total attenuation coefficient at a given height $r$ in the atmosphere, the cross-sections of each chemical species are interpolated using the values of $P$ and $T$ found at $r$ using the equation of hydrostatic equilibrium and the prescribed $P$--$T$ profile. This yields a two-dimensional array $\mu_i (\lambda, r)$ for each species. In the present work, since each combination of values of $\phi$ and $\theta$ may have a different $P$--$T$ profile, the cross-sections must be interpolated for each $(\phi,\theta)$ pair, yielding a four-dimensional array $\mu_i (\lambda, r, \theta, \phi)$. The attenuation coefficient for all chemical species acting as absorbers in the model is obtained by summing the attenuation coefficients for each individual species.
 
 We also include collision-induced absorption (CIA) due to H$_2$-H$_2$ and H$_2$-He \citep{Richard2012}. The attenuation coefficient due to CIA is given by the expression
 \begin{equation}
     \mu_{\rm{CIA}} = X_{\rm{H}_2} n_{\rm tot}^2 [ X_{\rm{H}_2} \sigma_{\rm{H}_2\rm{-H}_2} (\lambda,T) + X_{\rm{He}} \sigma_{\rm{H}_2\rm{-He}} (\lambda,T) ]
 \end{equation}
 where $\sigma_{\rm{H}_2\rm{-H}_2}$ and $\sigma_{\rm{H}_2\rm{-He}}$ are the H$_2$-H$_2$ and H$_2$-He cross-sections respectively. We note that unlike the cross-sections for chemical species described above, which have units of m$^2$, the CIA cross-sections have units of m$^5$.

\subsection{A Multidimensional Parametric Temperature Profile} \label{subsec:temp}

Our radiative transfer model is capable of taking any 3D temperature structure as its input. The model can therefore be used for post-processing GCMs to create transmission spectra. However, given that a single GCM typically takes at least several days to compute, such a model is not appropriate for atmospheric retrieval, which can require $\gtrsim 10^6$ model computations in order to analyse a single spectrum \citep{Fortney2021}. The temperature profile must therefore be simplified to enable rapid computation. The simplest approach is to consider an entirely isothermal temperature profile when carrying out retrievals \citep[e.g.][]{Waldmann2015,Zhang2019}. The analytic temperature profile described in \citet{Guillot2010} has been incorporated into retrieval algorithms \citep[e.g.,][]{Benneke2012,Line2013}, and the parametric temperature profile described in \citet{Madhu2009} has also been employed in numerous retrieval frameworks \citep[e.g.,][]{Pinhas2018,Blecic2022}. While the latter two approaches result in a temperature profile that varies with height in the atmosphere, these profiles do not vary with longitude or latitude.

\textsc{Aura-3D} incorporates several different parametric prescriptions for $P$-$T$ profiles that include temperature variations in multiple dimensions. This enables fast, flexible model computation while also allowing for varying levels of complexity in the temperature structure. We present three possible prescriptions: one in which temperature varies with height in the atmosphere but not with $\theta$, one in which temperature varies with $\theta$ but not with height, and one in which temperature varies with height as well as with $\theta$ and $\phi$.

\subsubsection{Temperature varying with height only} \label{subsubsec:1dpt}

In the case where the temperature profile does not vary with $\theta$ or $\phi$, we adopt the 1D parameteric $P$--$T$ profile from \citet{Madhu2009}. The atmosphere is divided into three layers, defined by $P_{1,2,3}$. The deepest layer is isothermal while the upper layers have thermal gradients controlled by $\alpha_{1,2}$. The full profile is defined as: 
\begin{align}
    P &= P_0 e^{\alpha_1\sqrt{T-T_0}}, & P_0 < P < P_1, \nonumber \\
    P &= P_2 e^{\alpha_2\sqrt{T-T_2}}, & P_1 < P < P_3, \\
    T &= T_3, & P > P_3, \nonumber
\end{align}
where $(P_0,T_0)$ defines the conditions at the top of the atmosphere. This can be recast to yield expressions for $T(P)$:
\begin{align}
    T(P) =
    \begin{cases}
    T_0 + \Bigg( \dfrac{\ln (P/P_0)}{\alpha_1} \Bigg)^2, & P_0 < P < P_1, \\
    T_2 + \Bigg( \dfrac{\ln (P/P_2)}{\alpha_2} \Bigg)^2, & P_1 < P < P_3, \\
    T_2 + \Bigg( \dfrac{\ln (P_3/P_2)}{\alpha_2} \Bigg)^2, & P > P_3.
    \end{cases}
\end{align}

This $P$--$T$ profile was developed to be capable of emulating observed temperature profiles of solar system planets as well as profiles from self-consistent exoplanet atmosphere models. This flexibility, along with a rapid computation time, makes it an ideal prescription for retrievals. Our new profile presented in section \ref{subsubsec:3dpt} applies the same philosophy to 3D models in order to extend our retrieval framework to incorporate multidimensional effects.

\subsubsection{Temperature varying with $\theta$ only} \label{subsubsec:thetapt}

This model provides a simple means of incorporating day-night temperature variations across the terminator region.  It is described by four main parameters. The extent of the terminator region over which the temperature varies is fixed by an angle $\beta$, similarly to \citet{Caldas2019}. The other three parameters are $T_{\rm term}$, the temperature in the middle of the terminator region ($\theta=\pi/2$), $T_{\rm day}$, the temperature at the dayside end of the terminator ($\theta=\pi/2 - \beta/2$), and $T_{\rm night}$, the temperature at the nightside end of the terminator ($\theta=\pi/2 + \beta/2$). The temperature at intermediate values of $\theta$ is found by linearly interpolating between the two appropriate temperatures. Outside of the fixed terminator region, the temperature remains uniform at either $T_{\rm day}$ or $T_{\rm night}$. This approach differs from the formalism described in \citet{Lacy2020} as it uses three different temperatures rather than just two ($T_{\rm day}$ and $T_{\rm night}$). This allows for the temperature gradient between the terminator and the dayside to be different to the gradient between the terminator and the nightside, which has been shown to be possible from GCMs, such as the model of HD~209458b presented in \citet{Fortney2010}.

\subsubsection{Temperature varying in three dimensions} \label{subsubsec:3dpt}

Our new temperature parameterisation generalises the 1D parametric $P$--$T$ profile described in Section \ref{subsubsec:1dpt} to allow for temperature variations in 3D. In order to achieve this, we compute three separate $P$--$T$ profiles located at three different values of $\theta$. We calculate one profile at $\theta = \pi/2$, the exact centre of the terminator, as well as profiles at the edges of the terminator which are closest to the day- and nightsides of the planet.  We label the dayside profile $T_d(P)$, the nightside profile $T_n(P)$, and the terminator profile $T_t(P)$. The terminator edges are located at $\theta = \pi/2 \pm \beta/2$, where $\beta$ is defined as in the previous section. At a given $(P,\theta)$ where $\theta$ is intermediate between the two terminator edges, the temperature is obtained by linear interpolation in $\theta$ between the two appropriate profiles:

\begin{align}
    T(P,\theta, \phi) = G(\phi)
    \begin{cases}
    \dfrac{2\theta - (\pi - \beta)}{\beta} T_t(P) \\
    + \dfrac{\pi - 2\theta}{\beta} T_d(P), & \dfrac{\pi - \beta}{2} < \theta < \pi/2, \\
    \dfrac{(\pi + \beta) - 2\theta}{\beta} T_t(P) \\
    + \dfrac{2\theta - \pi}{\beta} T_n(P), & \pi/2 < \theta < \dfrac{\pi + \beta}{2}.
    \end{cases}
\end{align}
where $G(\phi)$ is a generic function that can be used to describe the $\phi$-dependence of the temperature profile. For the purposes of this study we are mostly interested in exploring day-night temperature contrasts, and so we fix $G(\phi)=1$, however this can easily be modified to include $\phi$-dependence using a formalism similar to the patchy cloud approach described in Section \ref{subsubsec:clouds}.

Assuming that the $P$--$T$ profile is continuous throughout, it can be described using 6 free parameters. This suggests that the above parameterisation should require a total of 18 parameters. However, temperature variations with longitude/latitude are typically negligible in the deeper atmosphere \citep[e.g.,][]{Showman2008}, and so we can assign common values of $P_3$ and $T_3$ for each of the profiles, reducing the total number of free parameters to 14. A demonstration of the efficacy of this parameterisation is presented in Section \ref{subsec:pt_comparison}.

\subsection{Inhomogeneous chemistry}

Chemical inhomogeneities are likely to persist in a wide range of planets, and will therefore be important to consider when analysing upcoming data. Of particular importance will be day-night variations in chemical abundances. These variations can occur in ultra-hot Jupiters, planets with equilibrium temperatures in excess of 2000~K on which molecules such as H$_2$ and H$_2$O can be thermally dissociated on the dayside \citep{Lothringer2018,Parmentier2018,Venot2020}. They may also be present on warm to hot Neptunes, since their atmospheric composition varies strongly as a function of temperature and bulk atmospheric properties such as metallicity and C/O ratio \citep{Moses2013}. High-metallicity atmospheres for these planets are known to exhibit strong day-night temperature contrasts \citep{Lewis2010}, which could lead to a scenario in which the atmospheric composition could vary strongly between the day- and nightsides of the planet.

Our model can incorporate chemical compositions which vary in three dimensions. This extension requires the expansion of the number density of each species $n_i$, as defined in equation \ref{eq:opac} from a single number to a three-dimensional array $n_i(r,\theta,\phi)$. For the purpose of this study, we consider a simplified case in which two abundances are specified for each chemical species: a 'dayside' abundance ($\theta < \pi/2$) and a 'nightside' abundance ($\theta > \pi/2$). The effect of day-night chemical inhomogeneities on transmission spectra are explored in Section \ref{subsec:chem}. Considerations for retrievals with inhomogeneous chemical abundances are reserved for a future study (see section \ref{subsec:future} for further discussion).

\subsection{Additional physical effects}

Our model incorporates a wide range of other atmospheric properties. These are discussed in more detail in previous works \citep[e.g.,][]{Pinhas2018,Welbanks2021} and are briefly described here.

\subsubsection{Clouds and hazes} \label{subsubsec:clouds}

Our model includes a prescription for inhomogeneous cloud cover adapted from \citet{Welbanks2021}, which divides the atmosphere into (up to) four azimuthal slices. These slices can have one of four properties: (1) cloud/haze free, (2) covered by a grey cloud deck, (3) covered by hazes, or (4) covered by a grey cloud deck with hazes above the deck. The  grey cloud deck is essentially opaque \citep{Fortney2005} and is therefore implemented by setting the optical depth to infinity for all pressures larger than the cloud-top pressure $P_{\rm cloud}$. Similarly to previous studies \citep{Line2016,MacDonald2017,Welbanks2021} we adopt the haze model described in \citet{LDE2008}, which gives a haze cross-section
\begin{equation}
    \sigma_{\lambda, \rm haze} = a \sigma_0 \Bigg( \dfrac{\lambda}{\lambda_0} \Bigg)^{\gamma},
\end{equation}
where $a$ is the Rayleigh enhancement factor, $\gamma$ is the scattering slope, and $\sigma_0 = 5.31 \times 10^{-31}\,$m$^2$ is the cross-section due to H$_2$ Rayleigh scattering at a reference wavelength $\lambda_0 = 3.5 \times 10^{-7}\,$m \citep{Dalgarno1962}. The attenuation coefficient due to hazes, $\mu_{\lambda, \rm haze}$ is therefore given by the following expression:
\begin{equation}
    \mu_{\lambda, \rm haze} (\lambda, P, T) = X_{{\rm H}_{2}} n_{\rm tot} (P,T) \sigma_{\lambda, \rm haze},
\end{equation}
where $X_{{\rm H}_{2}}$ is the H$_2$ abundance and $n_{\rm tot} (P,T)$ is the total number density.

The above prescription can easily be simplified to yield the cloud model from \citet{MacDonald2017} by considering only slices with properties (1) and (4) or the model from \cite{Line2016} by considering only properties (1) and (2).

\subsubsection{Stellar heterogeneity}

In order to account for the effect of star spots and faculae on transmission spectra, we follow the approach of \citet{Pinhas2018} and incorporate the treatment of stellar heterogeneity from \citet{Rackham2017} into our atmospheric forward model. In cases where stellar heterogeneity is considered, the expression for the observed transit depth, $\Delta_{\lambda, \rm obs}$, is
\begin{equation}
    \Delta_{\lambda, \rm obs} = \Delta_{\lambda} \mathcal{E}_{\rm het},
\end{equation}
where $\Delta_{\lambda}$ is defined in Equation \ref{eq:td} and $\mathcal{E}_{\rm het}$ is a perturbative term describing stellar contamination. $\mathcal{E}_{\rm het}$ is defined as
\begin{equation}
    \mathcal{E}_{\rm het} = \left( \left( 1 - \frac{\mathcal{S}_u}{\mathcal{S}_o} \right) f_{\rm het} \right)^{-1},
\end{equation}
where $\mathcal{S}_u$ and $\mathcal{S}_o$ are the average spectral energy distributions of the unocculted and occulted regions of the stellar surface, and $f_{\rm het}$ is the areal fraction of the projected stellar disk that is covered with cool spots and/or hot faculae. The spectral components of the star are computed by interpolating from the PHOENIX grid of stellar atmospheric models \citep{Husser2013} for stellar effective temperatures exceeding 2300~K. For stars cooler than 2300~K we instead interpolate between the DRIFT-PHOENIX model grid \citep{Witte2011}, which enables consideration of stars with temperatures down to 1000~K.

\subsubsection{Forward scattering and refraction}

Our forward model incorporates the analytic prescriptions for forward scattering and refraction presented by \citet{Robinson2017}. The optical depth is typically calculated by integrating along $\rm{d} \tau_{\lambda} = \mu_{\lambda}$d$s$ (see equation \ref{eq:tau}). When forward scattering is included, the optical depth is modified to become
\begin{equation}
    \textnormal{d} \tau_{\rm eff}  = \textnormal{d} \tau_{\lambda} ( 1 - f_{\rm scat} \Tilde{\omega} ),
\end{equation}
where $f_{\rm scat}$ is the correction factor for forward scattering and $\Tilde{\omega}$ is the forward scattering albedo. Refraction is implemented by calculating $P_{\rm max}$, the pressure at which the effect of refraction is sufficient to cause a ray of light coming from one end of the planet to bend off the far limit of the star \citep{Robinson2017}. The optical depth at pressures greater than $P_{\rm max}$ is set to infinity.

The effects of forward scattering and refraction on transmission spectra were investigated by \citet{Robinson2017} for Jupiter-like atmospheres and by \citet{Welbanks2021} for mini-Neptunes. In both cases the effects were shown to be small compared to effects such as Rayleigh scattering and collision-induced absorption.

\subsection{Synthetic Data and Statistical Inference}

For this study we generate synthetic JWST observations and transmission spectra using PANDEXO \citep{Batalha2017}. For all simulated observations, we assume a noise floor of 5 ppm, and a saturation limit of 80\% full well. For atmospheric retrieval, we follow the binning strategy of \citet{Pinhas2018} in order to compare the high-resolution forward model with the synthetic data.

Parameter estimation is carried out using PyMultinest \citep{Buchner2014}, a Python implementation of the Nested Sampling algorithm \citep{Skilling2004,Feroz2008,Feroz2009}. For each parameter $\theta$ we aim to find the posterior probability distribution $p(\theta|d)$ given the data $d$ being analysed:
\begin{equation}
    p(\theta|d) = \frac{\mathcal{L} p(\theta)}{\mathcal{Z}},
\end{equation}
where $\mathcal{L}=p(d|\theta)$ is the likelihood, $p(\theta)$ is the prior and $\mathcal{Z} = p(d)$ is the Bayesian evidence, which is not required for parameter estimation but can be used for model comparison. We assume independently distributed Gaussian errors for each of the spectral data points, meaning the likelihood is defined as
\begin{equation}
    \mathcal{L} = \mathcal{L}_0 \exp \bigg( - \frac{\chi^2}{2} \bigg),
    \label{eq:like}
\end{equation}
\begin{equation}
    \chi^2 = \sum_i \frac{(\hat{y}_i-\bar{y}_i)^2}{\sigma_i^2},
    \label{eq:chi}
\end{equation}
where $\bar{y}_i$ and $\sigma_i$ are the mean and standard deviation of the observed data point $i$, and $\hat{y}_i$ is the value of the model data point $i$. The Nested Sampling algorithm samples the prior space in order to calculate the evidence $\mathcal{Z}$, and in the process of doing so computes the likelihood $\mathcal{L}$ which can be used to estimate the posterior distribution.

\subsection{Model validation}

\begin{figure*}
\includegraphics[width=\linewidth]{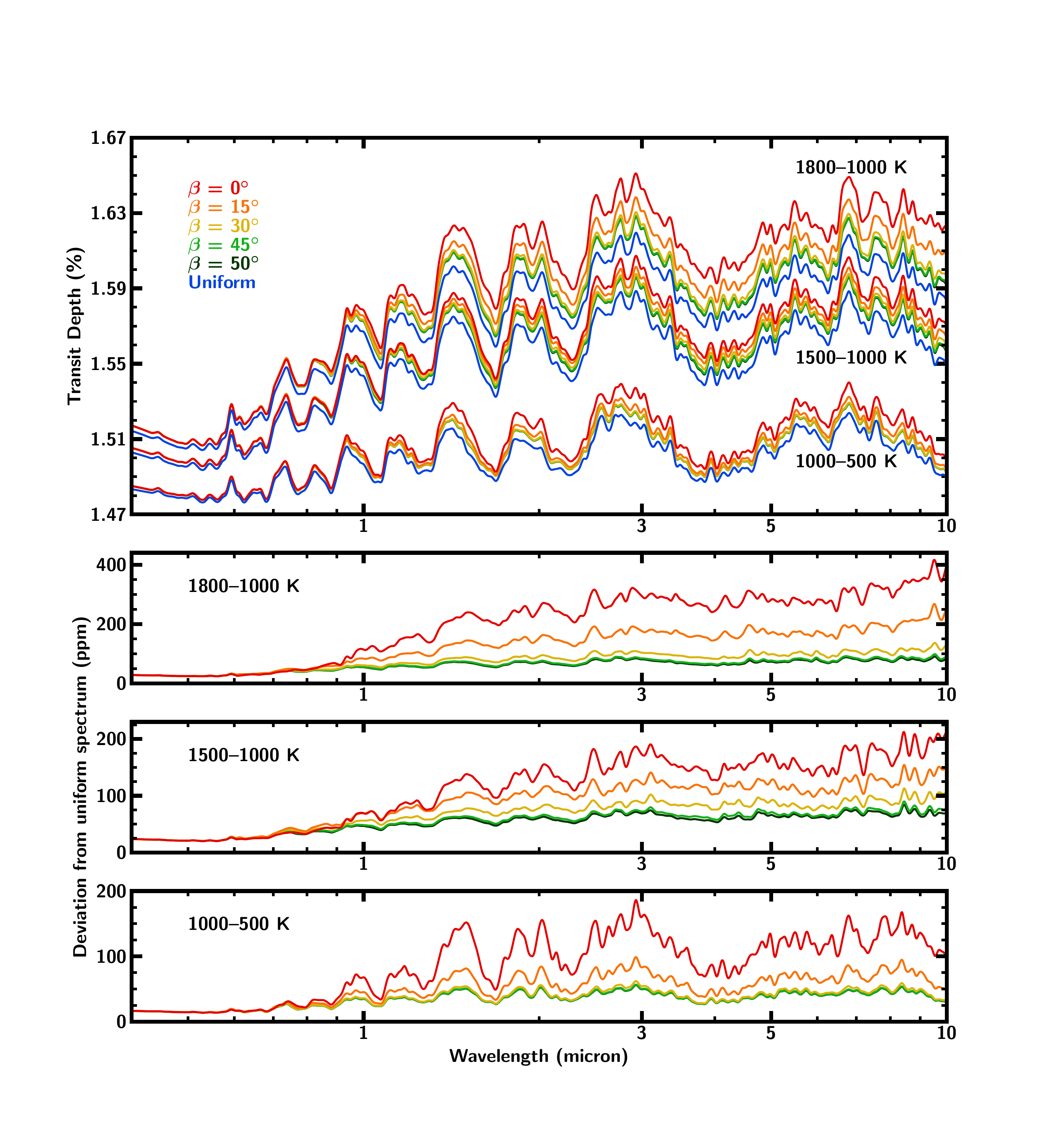}
    \caption{\textit{Upper panel:} Model transmission spectra of the canonical hot Jupiter HD~209458b with varying temperatures and transition angles ($\beta$). For these models, we adopt the parametric temperature profile described in \citet{Caldas2019}. We obtain the same effects as described in their paper: namely, that the transit depth decreases as $\beta$ increases, and the divergence between models increases at larger wavelengths. \textit{Lower panels:} Difference in ppm between spectra with the non-uniform temperature profiles described above and uniform spectra with averaged temperature profiles.}
    \label{fig:caldas_plot}
\end{figure*}

In order to verify that our 3D forward model is accurate, we reproduce effects reported in previous works that have incorporated a 3D transmission spectrum model. We consider the set of model transmission spectra of the hot Jupiter HD~209458b presented in Figure 15 of \citet{Caldas2019}. We adopt the parametric $P$--$T$ profile used in that model, which consists of a two-layer vertical profile, with a constant temperature in the deep atmosphere and a variable temperature from day- to nightside in the upper atmosphere. We produce a set of models with the same temperature structures and transition angles ($\beta$) as in \citet{Caldas2019}, as shown in Figure \ref{fig:caldas_plot}. Our models show the same dependencies on temperature and transition angle as the previous work, with higher transit depths found for lower values of $\beta$.

While we have adopted the temperature profile described in \citet{Caldas2019} for the purposes of validating our model, we note that this prescription is simplified compared to our parametric $P$--$T$ profile described in Section \ref{subsubsec:3dpt}. Unlike our parametric profile, their parameterisation is not intended to be capable of matching temperature structures from GCMs (see Section \ref{subsec:pt_comparison}).

\begin{figure*}
\includegraphics[width=0.95\textwidth]{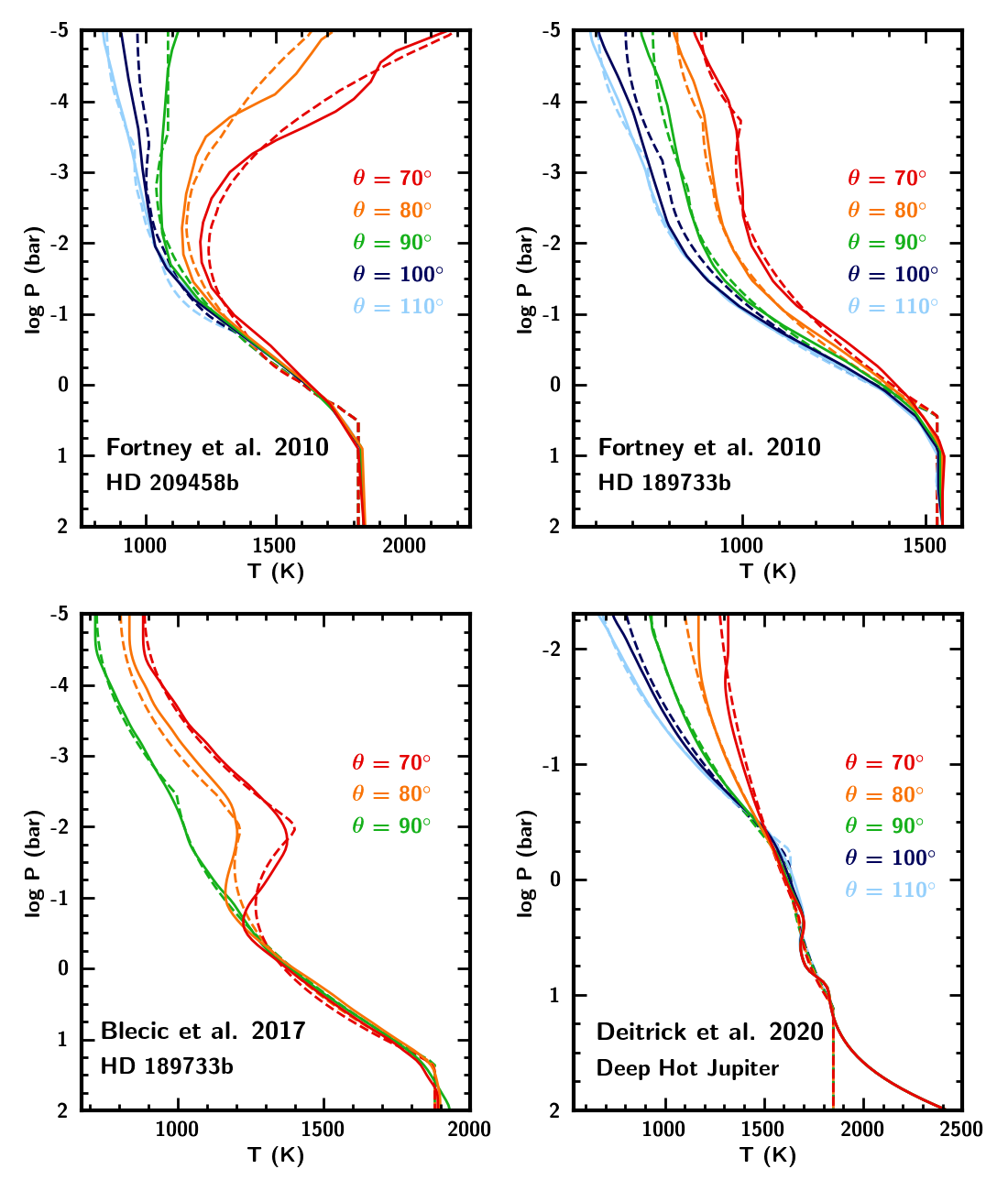}
    \caption{Comparison between azimuthally-averaged $P$--$T$ profiles from the terminator regions of a range of GCMs \citep{Fortney2010,Blecic2017,Deitrick2020} and parametric profiles as described in Section \ref{subsec:temp}. The GCM profiles are shown as solid lines, and parametric fits are shown as dashed lines. In each case the profiles at $\theta = 70^{\circ},\,90^{\circ}$ and $110^{\circ}$ are fit using a modified form of the parameterisation from \citet{Madhu2009}, with the parameters $P_3$ and $T_3$ shared between each profile. The profiles at $\theta = 80^{\circ}$ and $100^{\circ}$ are calculated by linear interpolation between the nearest two profiles. For the \citet{Blecic2017} model, we only show profiles for $\theta \leq 90^{\circ}$ since nightside profiles were not presented in that paper.}
    \label{fig:pt_fit}
\end{figure*}

\section{Results}
\label{section:results}

In this section we demonstrate that our 3D parametric $P$--$T$ profile can match azimuthally-averaged temperature structures from established GCMs. We subsequently explore the extent to which day-night temperature gradients and chemical inhomogeneity can affect resulting transmission spectra. We also perform 3D retrievals on a number of synthetic datasets, demonstrating \textsc{Aura-3D}'s capabilities to constrain temperature profiles which vary across the terminator and to overcome biases in retrieved abundances that can arise from the assumption of a 1D temperature profile.

\subsection{Comparison of parametric profile with GCM temperature profiles} \label{subsec:pt_comparison}

\begin{figure*}[t]
\centering
\begin{center}$
\begin{array}{cc}
\includegraphics[width=0.47\textwidth,trim={2.5cm 0 1.5cm 0},clip]{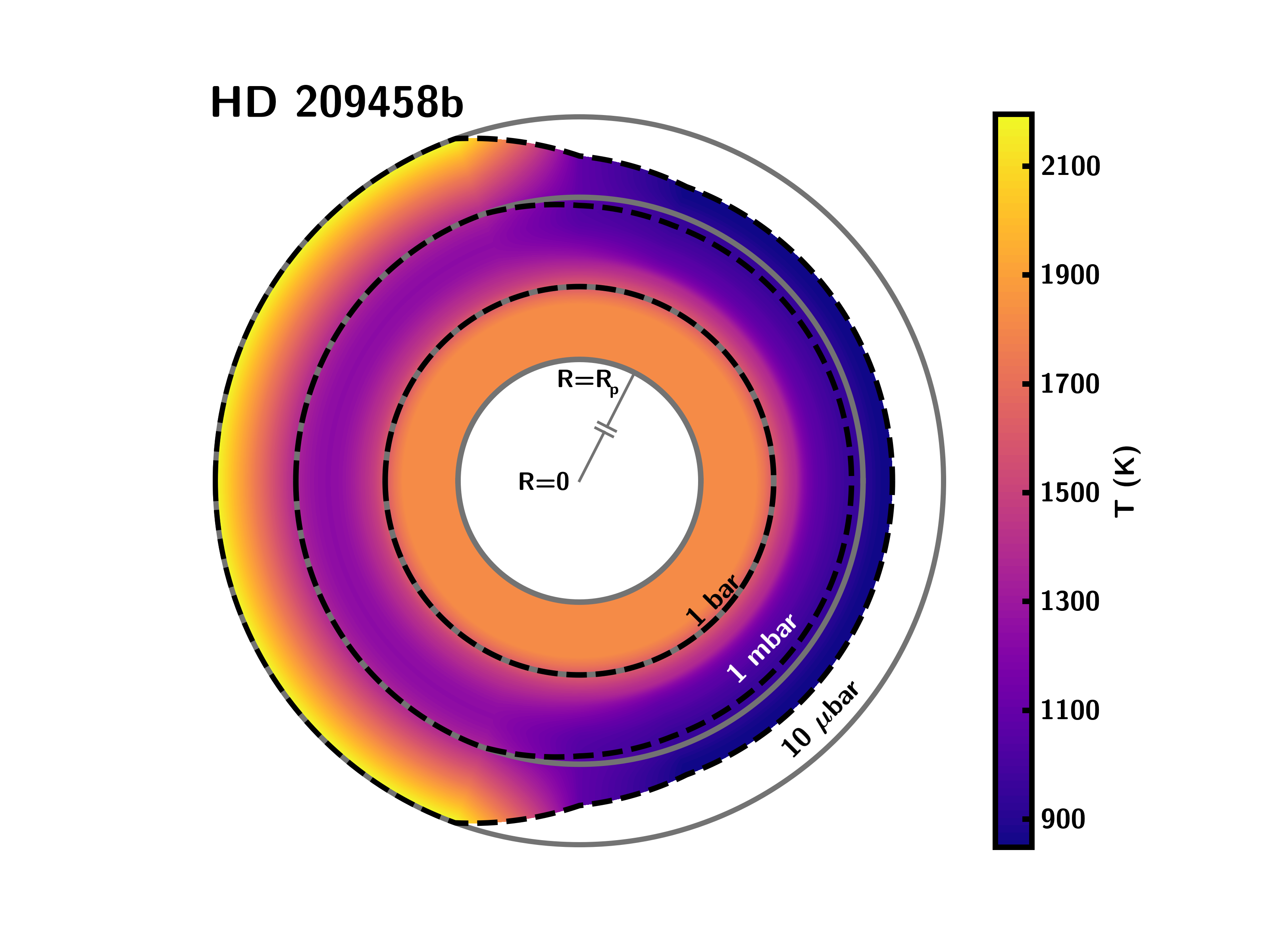}
\includegraphics[width=0.47\textwidth,trim={2.5cm 0 1.5cm 0},clip]{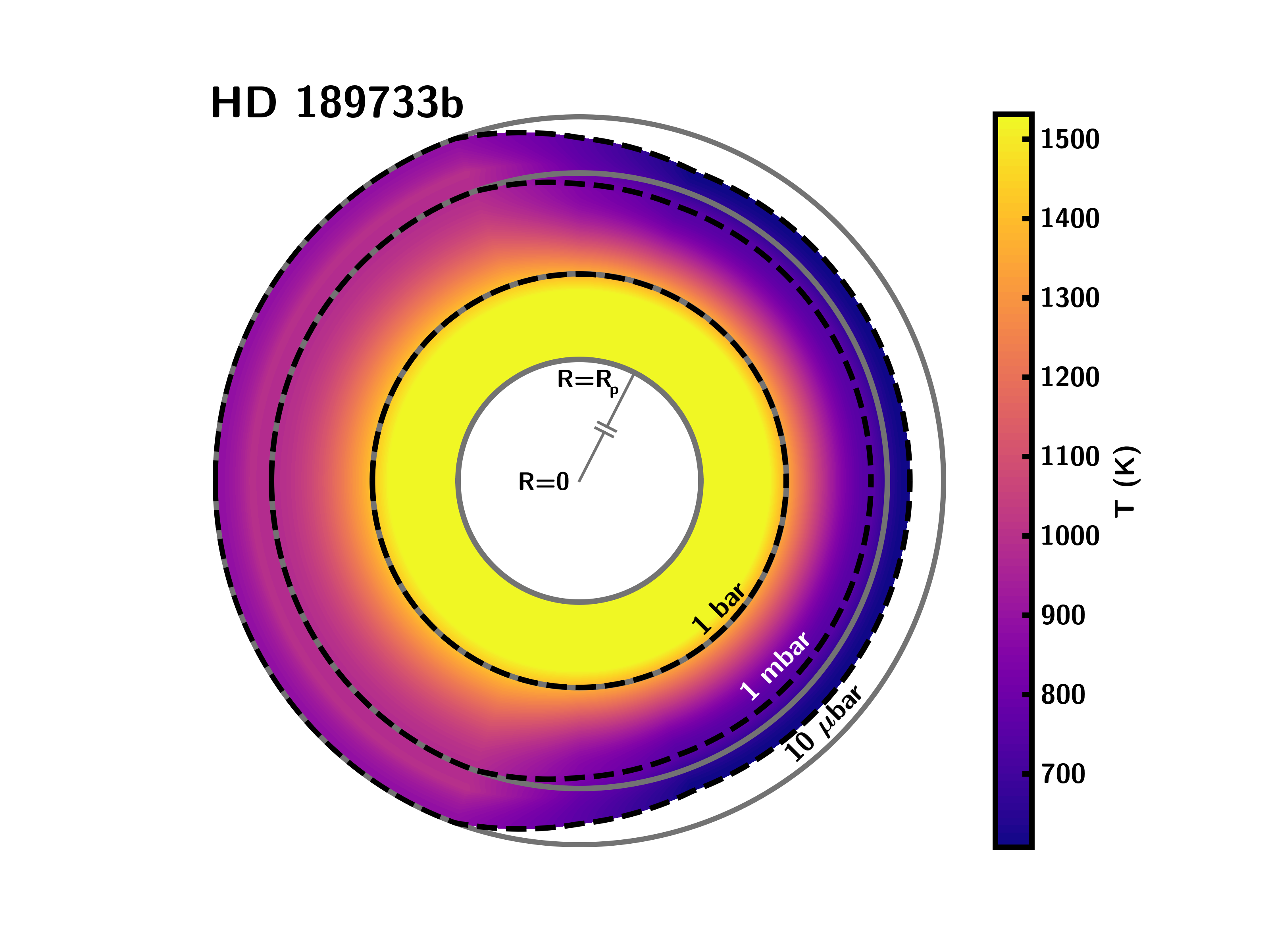}
\end{array}$
\end{center}
    \caption{Parametric fits to the 3D $P$--$T$ profiles presented in \citet{Fortney2010} showing a slice at $\phi=0$ (through the north-south polar plane). As in Figure \ref{fig:pt_fit}, the profiles at $\theta = 70^{\circ},\,90^{\circ}$ and $110^{\circ}$ follow our parametric prescription while the profiles all other values of $\theta$ are calculated by interpolation. For $\theta < 70^{\circ}$ and $\theta > 110^{\circ}$ the temperature remains constant, set by the temperatures at $\theta = 70^{\circ}$ and $\theta = 110^{\circ}$ respectively. Note that the inner portion of the planet (below $R_p$) is not to scale with the atmosphere.}
    \label{fig:polar_pt}
\end{figure*}

We demonstrate that our 14-parameter prescription for the temperature profile is able to approximate a range of azimuthally-averaged temperature structures of several GCMs. Figure \ref{fig:pt_fit} shows temperature profiles from the terminator regions of several published 3D simulations of hot Jupiter atmospheres. We consider the models of HD~209458b and HD~189733b presented in \citet{Fortney2010}, the model of HD~189733b from \citet{Blecic2017}, and the Deep Hot Jupiter benchmark model from \citet{Deitrick2020}. For each model we show the azimuthally-averaged temperature profiles across the terminator region of the atmosphere. Since only dayside profiles are presented in \citet{Blecic2017} we only show the temperature profiles at $\theta = 70^{\circ},\,80^{\circ}$ and $90^{\circ}$ for this case. To fit these models we assume a value of $\beta = 40^{\circ}$. For each model, we therefore fit the temperature profiles at $\theta = 70^{\circ},\,90^{\circ}$ and $110^{\circ}$ with our 14-parameter prescription, with $T_3$ and $P_3$ shared between the profiles, using a Trust Region Reflective least-squares algorithm \citep{Sorenson1982}. The full temperature structure is subsequently calculated by linear interpolation between the nearest of those three profiles. Figure \ref{fig:pt_fit} includes profiles at $\theta = 80^{\circ}$ and $100^{\circ}$ in order to demonstrate the accuracy of the interpolated profiles. The full distribution of day-night temperature profiles for the models presented in \citet{Fortney2010} are shown in Figure \ref{fig:polar_pt}.

Our parametric $P$--$T$ prescription is able to fit all of the profiles shown in Figure \ref{fig:pt_fit} to a good degree of accuracy. The model of HD~209458b from \citep{Fortney2010} is a good demonstration of why interpolation between three temperature profiles is preferred to two. In this case, the difference between the temperature profile at $\theta=70^{\circ}$ and the profile at $\theta=90^{\circ}$ is much greater than the difference between the profiles at $\theta=90^{\circ}$ and $\theta=110^{\circ}$. If the parametric model only interpolated between the profiles at $\theta=70^{\circ}$ and $\theta=110^{\circ}$, then the interpolated $\theta=90^{\circ}$ would be much hotter than the profile from the GCM and would yield a worse fit. 

For the Deep Hot Jupiter case, we did not fit our profile to the deep atmosphere ($>$10 bar) since this is below the region of the atmosphere probed in transmission, hence the deviation between the GCM and parametric profiles at high pressure. For atmospheric retrieval, the goal is to quickly generate a large number of model spectra which can be compared to observed data. Current retrievals require the generation of $\gtrsim$10$^6$ forward models \citep{Fortney2021}. Generating a temperature structure from a GCM typically takes several days, rendering them unfeasible for retrievals. In contrast, a forward model using our parametric profile is computed in $\sim$0.6 seconds on a single core. Our parameterisation therefore enables the efficient exploration of a wide range of multidimensional temperature structures.

In order to confirm that our parametric $P$--$T$ profile can be used to generate transmission spectra closely matching those generated with a full 3D temperature structure, we produce a model spectrum from 0.5--5.5$\mu$m using the Deep Hot Jupiter temperature structure from \citet{Deitrick2020}, assuming a H/He-rich atmosphere with H$_2$O as the sole absorbing species. We compare this to a spectrum generated using our parametric fit shown in Figure \ref{fig:pt_fit}. The mean difference between the transit depths in the two spectra is 12.1~ppm, which is likely below the noise floor of most JWST instruments \citep{Greene2016}. In contrast, comparison between the full 3D model and a 1D best-fit to the globally-averaged temperature profile yields a difference of 65.5~ppm, indicating that our parameterisation provides a better match to the 3D spectrum than a 1D model. The differences between the spectra are more pronounced at longer wavelengths (see Figure \ref{fig:1d_3d_contrast_grid}).

\subsection{Effects of a day-night temperature gradient on atmospheric spectra} \label{subsec:mgrid}

We use our parametric temperature structure to explore how a day-night temperature gradient can lead to differences in resulting transmission spectra from 1D averaged models. The benefit of using the parametric model to investigate this is that a large number of spectra can be generated quickly for comparision, without the need to run a separate GCM for each temperature structure. In order to explore this effect we generate a grid of model hot Jupiter transmission spectra. We take the planetary and stellar parameters of the hot Jupiter HD~209458b \citep{Stassun2017}, and create models with a variety of temperature structures assuming solar H$_2$O abundance. In the middle of the terminator region ($\theta=90^{\circ}$), we vary the temperature at the top of the atmosphere in steps of 100~K from 1000~K to 1800~K, covering the range of typical hot Jupiters. We also vary the temperature contrast, $\Delta T$, in steps of 5~K, from 0~K to 700~K. This temperature contrast is applied to both the day- and nightsides, so that $T_d=T_t+\Delta T$ and $T_n=T_t-\Delta T$. We assume $\beta=40^{\circ}$, meaning the temperature varies between $\theta=70^{\circ}$ and $\theta=110^{\circ}$. The other parameters describing the temperature structure are held constant: the common temperature in the deepest layer of the atmosphere, set by $T_3$, remains at $T_t+800$~K in all cases. The values of $\log P_1$, $\log P_2$ and $\log P_3$ are fixed to -0.9, -1.0 and 1.4 respectively. This means that none of the temperature profiles have a thermal inversion, since $P_1 > P_2$. 

\begin{figure*}
\includegraphics[width=\linewidth]{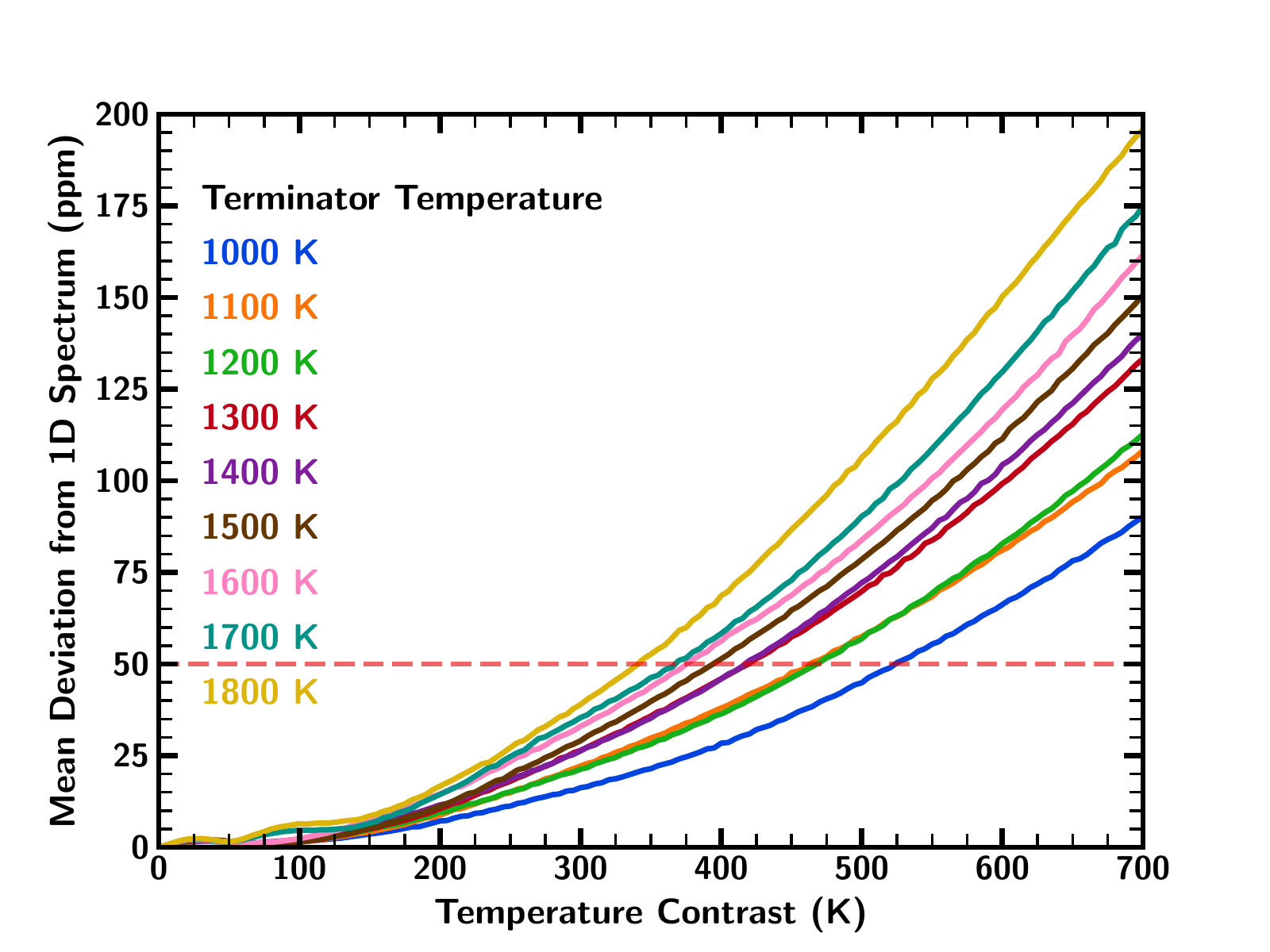}
    \caption{Differences between simulated hot Jupiter spectra with a 1D thermally averaged temperature profile and spectra with different day-night temperature gradients. For a range of terminator temperatures $T_t$ (the temperature at the top of the atmosphere at $\theta=90^{\circ}$) we compute spectra with temperature contrasts $\Delta T=0$--700~K ($\Delta T = T_d-T_t=T_t-T_n$). In each case, the spectra are calculated at $R=5000$ and a wavelength range of 1--15$\,\mu$m and the mean difference between the two transit depths is calculated. The difference in transit depth is given in parts per million. The dashed red line at 50 ppm is an approximate indication of the expected precision that can readily be achieved by JWST observations of hot Jupiters.}
    \label{fig:1d_3d_contrast_grid}
\end{figure*}

\begin{figure*}
\includegraphics[width=\linewidth]{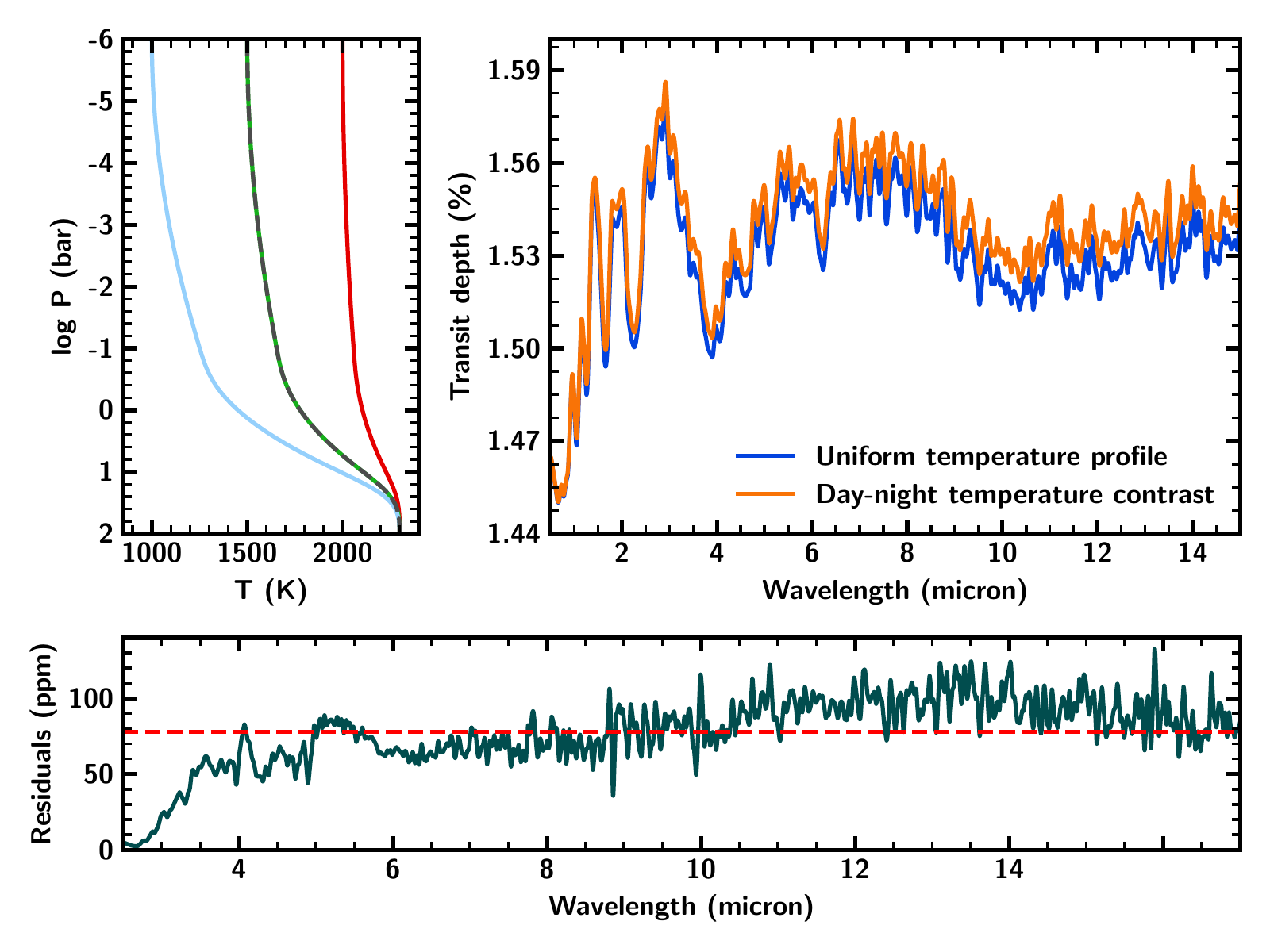}
    \caption{Comparison of two theoretical transmission spectra for a HD~209458b-like planet (top right). The orange spectrum uses a parametric temperature structure with a day-night temperature contrast, shown by the blue, green and red temperature profiles in the top left panel, whereas the blue spectrum uses a 1D averaged structure shown by the dashed grey temperature profile. The bottom panel shows the difference between the two spectra as a function of wavelength. We find that the spectra differ by 78 ppm on average, enough that high-quality JWST observations of this planet would require day-night temperature variations to be considered in retrievals.}
    \label{fig:spectrum_comparison}
\end{figure*}

We generate transmission spectra over a wavelength range of 0.5--15$\mu$m. This covers the wavelength range of several JWST instruments, including NIRSPEC FSS \& BOTS, NIRISS SOSS, NIRCAM GRISM and MIRI LRS. Each spectrum is computed at a moderately high resolution ($R=5000$). For each combination of $T_t$ and $\Delta T$, we calculate the mean difference with the 1D spectrum and present our full results in Figure \ref{fig:1d_3d_contrast_grid}. An example with $T_t=1500$~K, $\Delta T=500$~K is shown in Figure \ref{fig:spectrum_comparison}. These tables show that for a wide range of temperature contrasts, the difference between a spectrum with a temperature gradient and one without is non-trivial. In general, the difference is more pronounced at higher temperatures and at higher temperature contrasts. For all models except $T_t=1000$~K, the difference between the spectra is $>$50~ppm for a temperature contrast of 500~K or more. This level of precision would be expected for many JWST transmission spectra of hot Jupiters, depending on factors such as observing time and the brightness of the host star. However, the difference between the spectra is not uniform with wavelength, with less of a noticeable difference at short wavelengths (see Figure \ref{fig:spectrum_comparison}. This suggests that for some instruments, such as NIRISS SOSS, the temperature gradient will have less of an impact on the observed spectrum, while the effect will be more important to consider when observing with instruments such as MIRI MRS and LRS. It is also important to note that while some HST Wide Field Camera 3 (WFC3) transmission spectra have also attained a level of precision of 50~ppm or better, these models indicate that WFC3 spectra will not be substantially affected by day-night temperature gradients since the difference between the spectra is small at the wavelengths probed by WFC3 (1.1--1.7$\,\mu$m).

\begin{figure*}
\includegraphics[width=\linewidth]{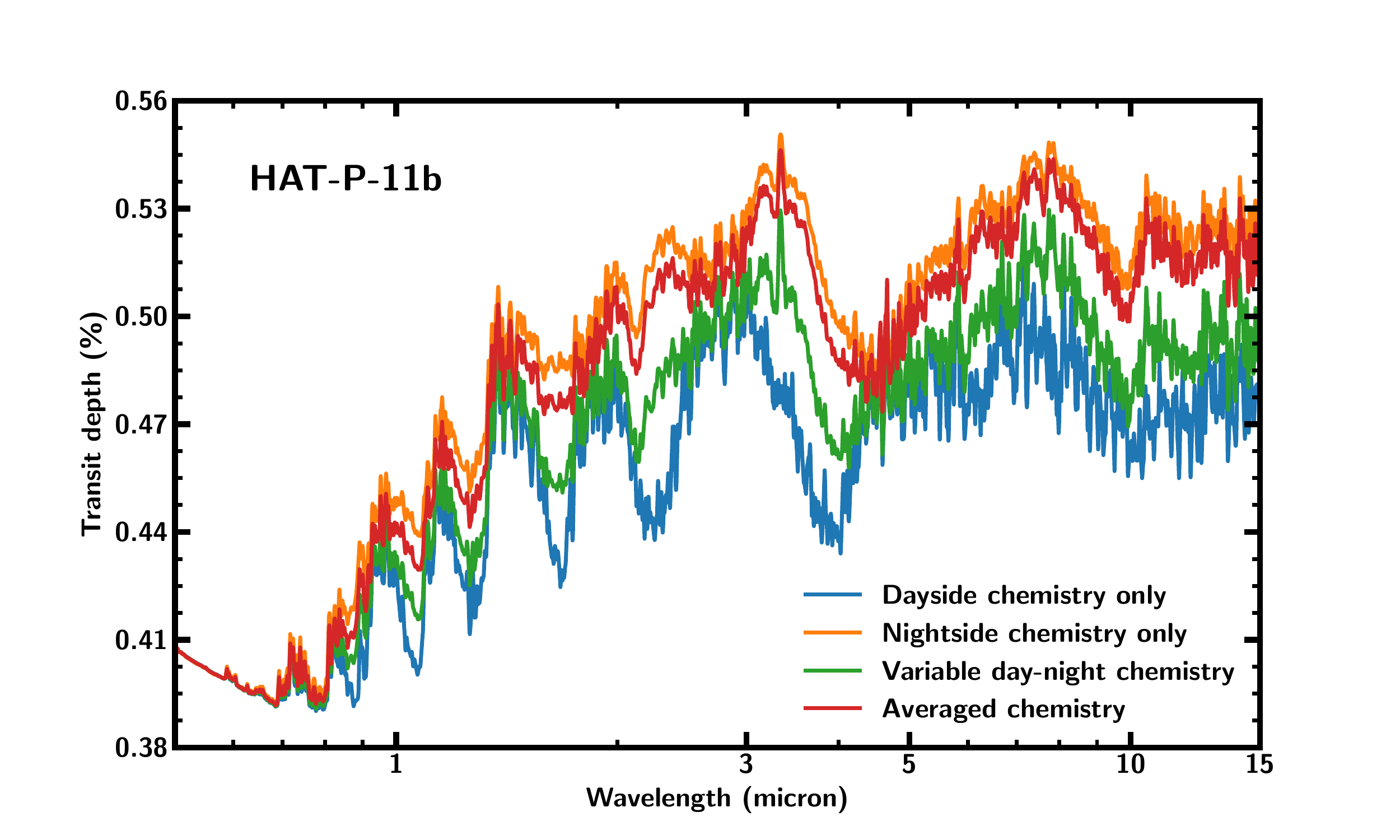}
    \caption{Synthetic spectra of the hot Neptune HAT-P-11b. Four different chemical compositions are considered: (1) `dayside' chemistry everywhere, (2) `nightside' chemistry everywhere, (3) variable chemistry between the day- and nightsides, and (4) averaged chemistry everywhere (see Table \ref{tab:abundances}). Model 2 generally has a higher transit depth than model 1 due to higher abundances of H$_2$O, CH$_4$ and NH$_3$ which are all prominent at these wavelengths. Model 3 generally lies between models 1 and 2, whereas model 4 lies closer to model 2, since the `nightside' molecules are also found on the dayside of the planet, where the scale height is larger due to the higher temperatures.}
    \label{fig:daynight_spectra}
\end{figure*}

\begin{figure*}[t]
\includegraphics[width=0.9\textwidth,trim={0cm 0cm 0 0cm},clip]{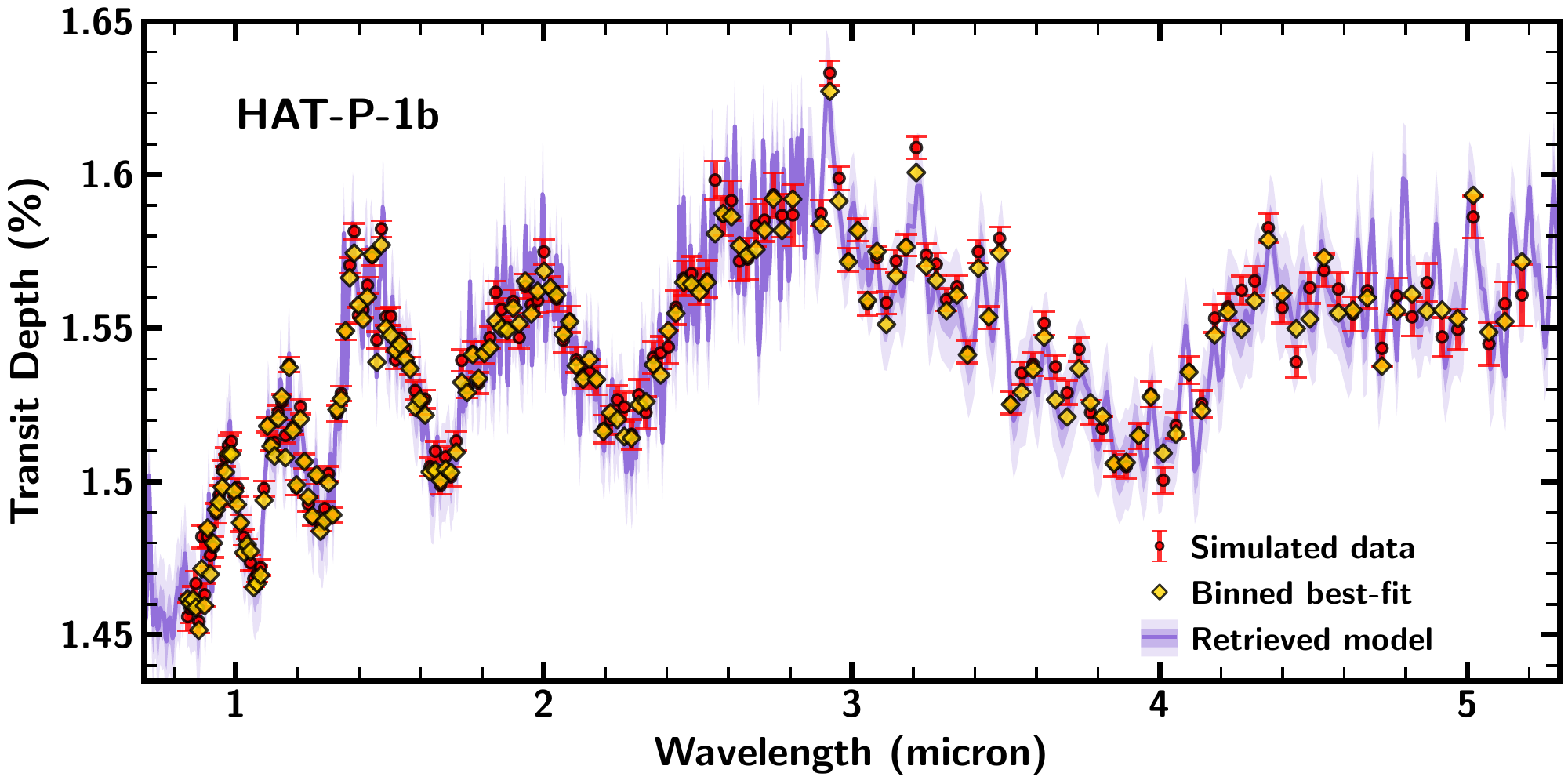}
    \caption{Retrieval of a synthetic JWST spectrum of the hot Jupiter HAT-P-1b. The red points represent the simulated data, and the purple line shows the best-fit model spectrum, with 1$\sigma$ and 2$\sigma$ confidence intervals shown by the dark and light shaded regions. The input and retrieved models both incorporate a multidimensional parametric $P$--$T$ profile. We find good agreement between the input data and the retrieved spectrum.}
    \label{fig:3d_par_ret_spec}
\end{figure*}

\begin{figure}
\includegraphics[width=\linewidth,trim={0.3cm 0cm 0 0cm},clip]{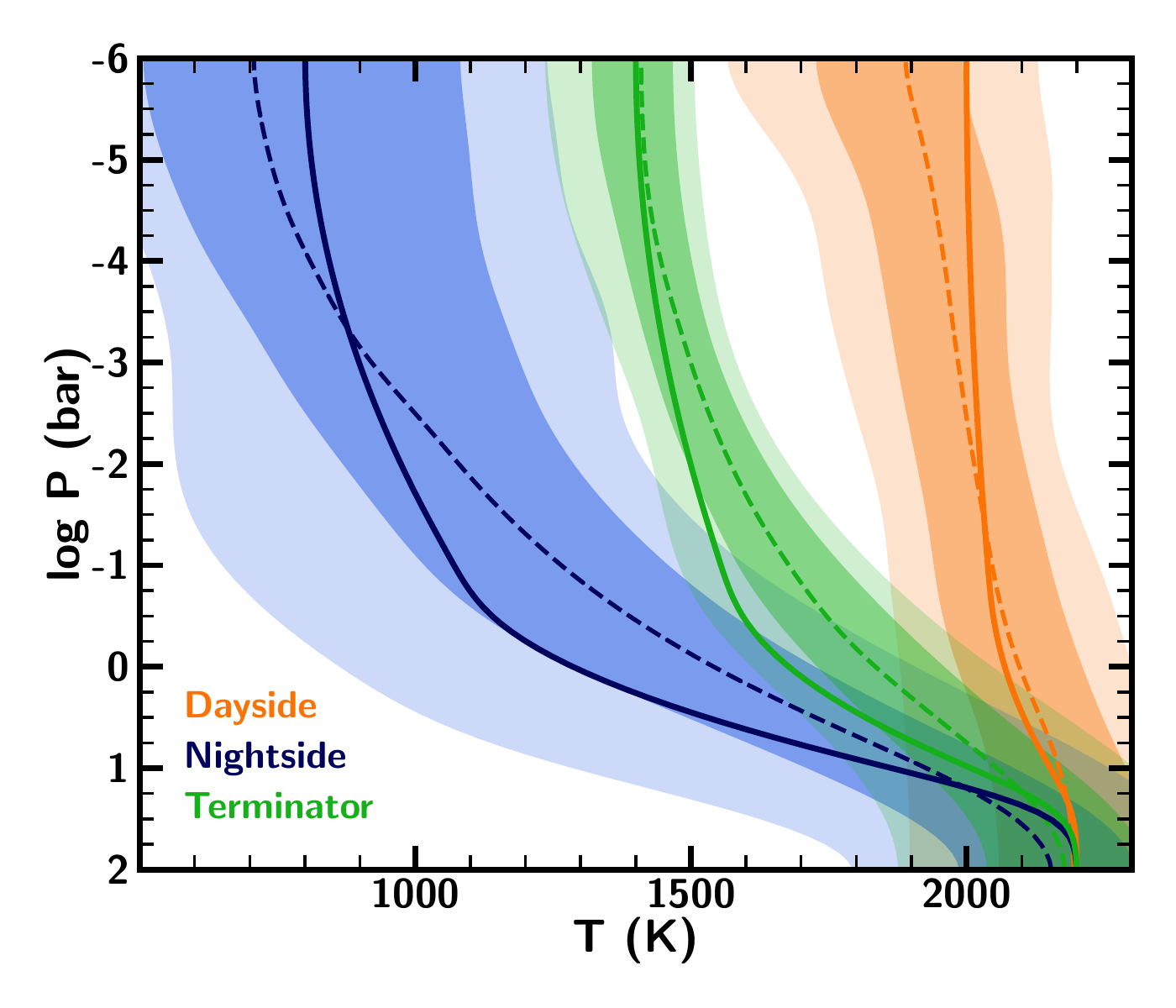}
    \caption{Input and retrieved multidimensional parametric temperature profiles of a hot Jupiter. The solid lines show the input dayside, nightside, and terminator temperature profiles used to generate the synthetic data. The dashed lines correspond to the median retrieved dayside, nightside and terminator temperature profiles respectively, with shaded regions corresponding to 1$\sigma$ and 2$\sigma$ confidence intervals. The retrieved temperature profiles are consistent with the input profiles, with almost all of the input temperature profiles contained within the retrieved 1$\sigma$ confidence intervals.}
    \label{fig:3d_par_ret}
\end{figure}

\begin{figure*}[t]
\centering
\begin{center}$
\begin{array}{cc}
\includegraphics[width=0.47\textwidth,trim={2.5cm 0 1.5cm 0},clip]{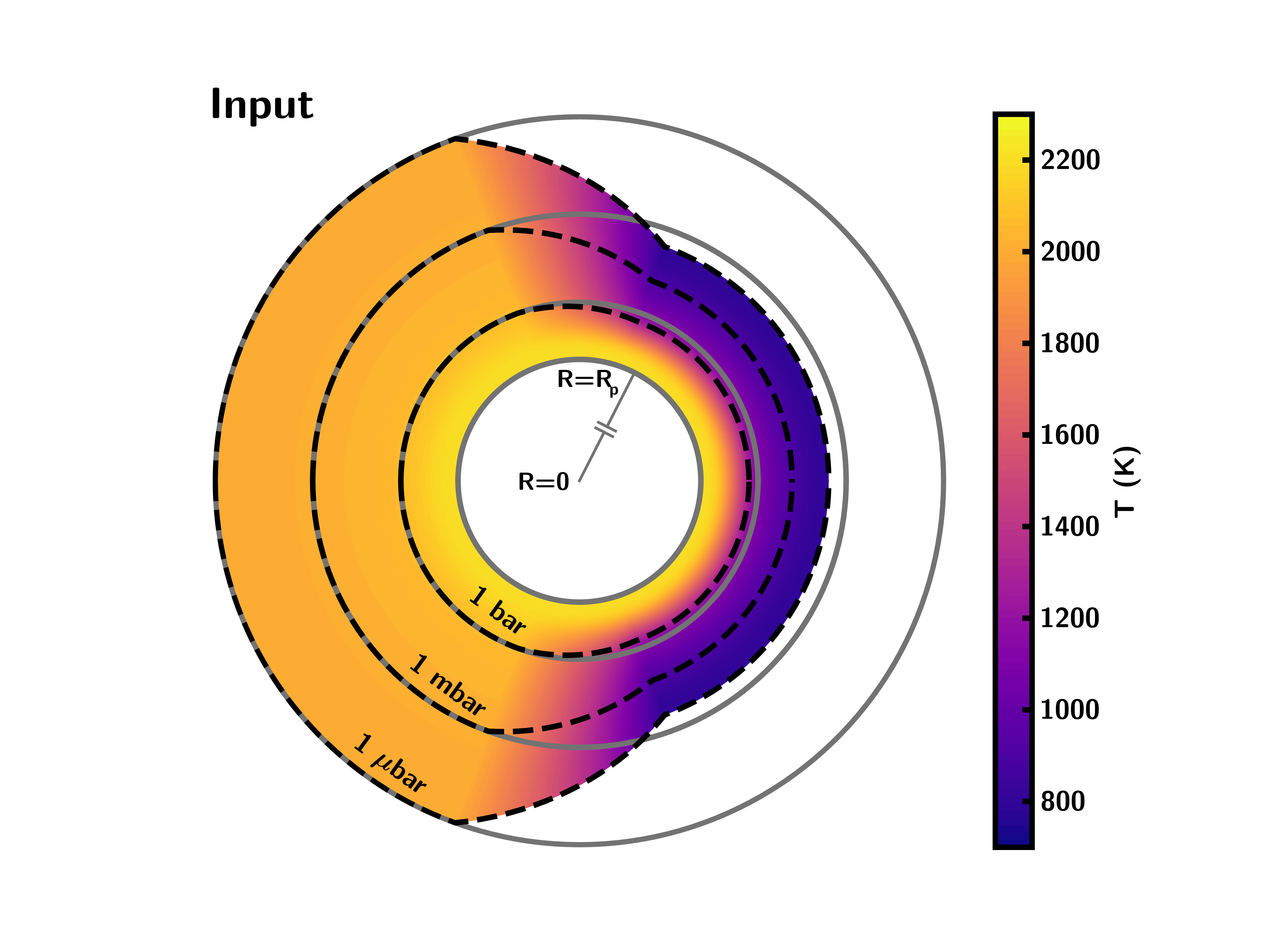}
\includegraphics[width=0.47\textwidth,trim={2.5cm 0 1.5cm 0},clip]{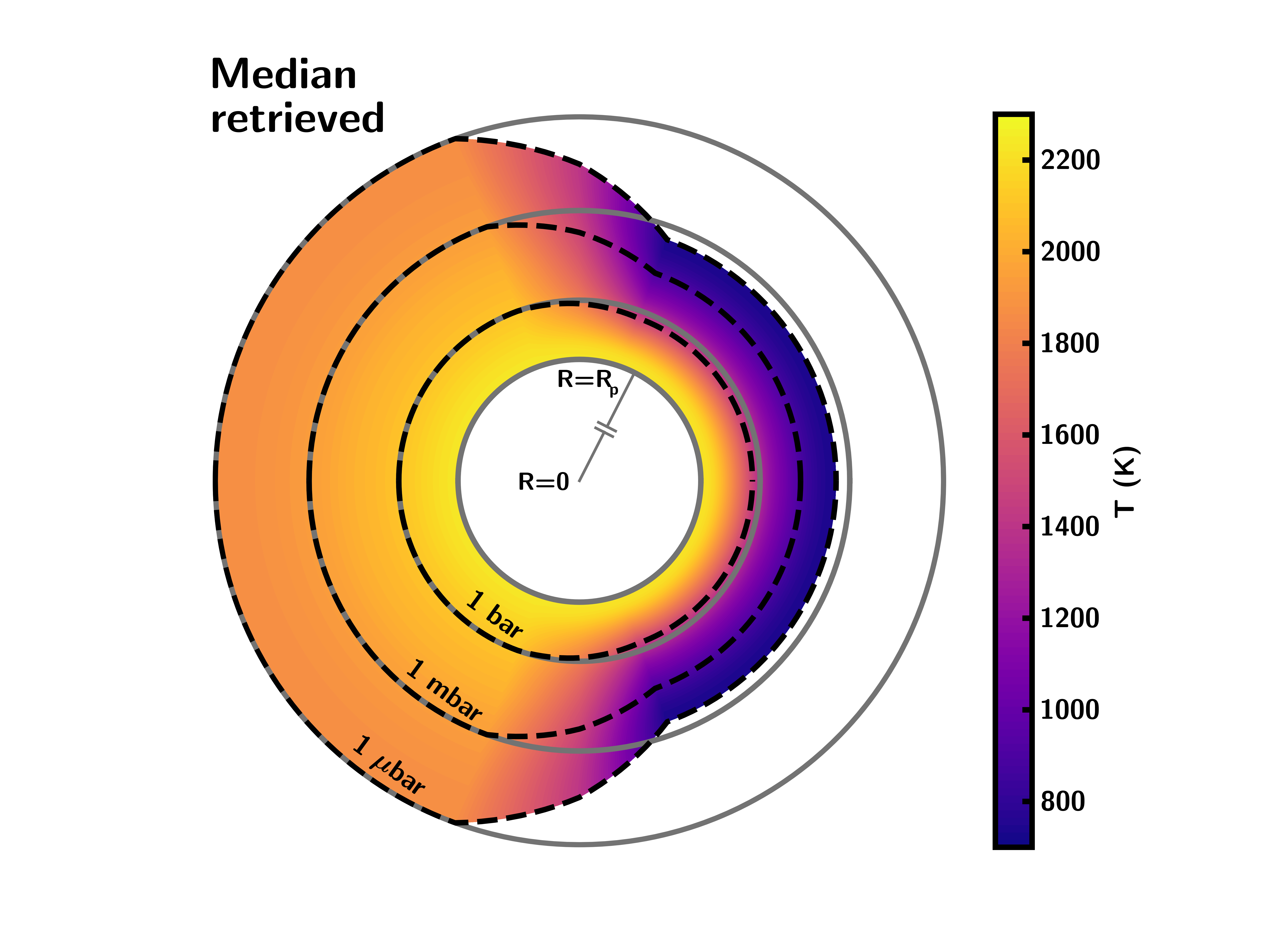}
\end{array}$
\end{center}
    \caption{Two-dimensional representations of the input and median retrieved parametric $P$--$T$ profiles for HAT-P-1b. The temperature structure is computed by interpolating between the temperature profiles shown in Figure \ref{fig:3d_par_ret}. Note that the inner portion of the planet (below $R_p$) is not to scale with the atmosphere.}
    \label{fig:polar_pt_ret}
\end{figure*}

\begin{figure}
\includegraphics[width=\linewidth,trim={0.3cm 0cm 0 0cm},clip]{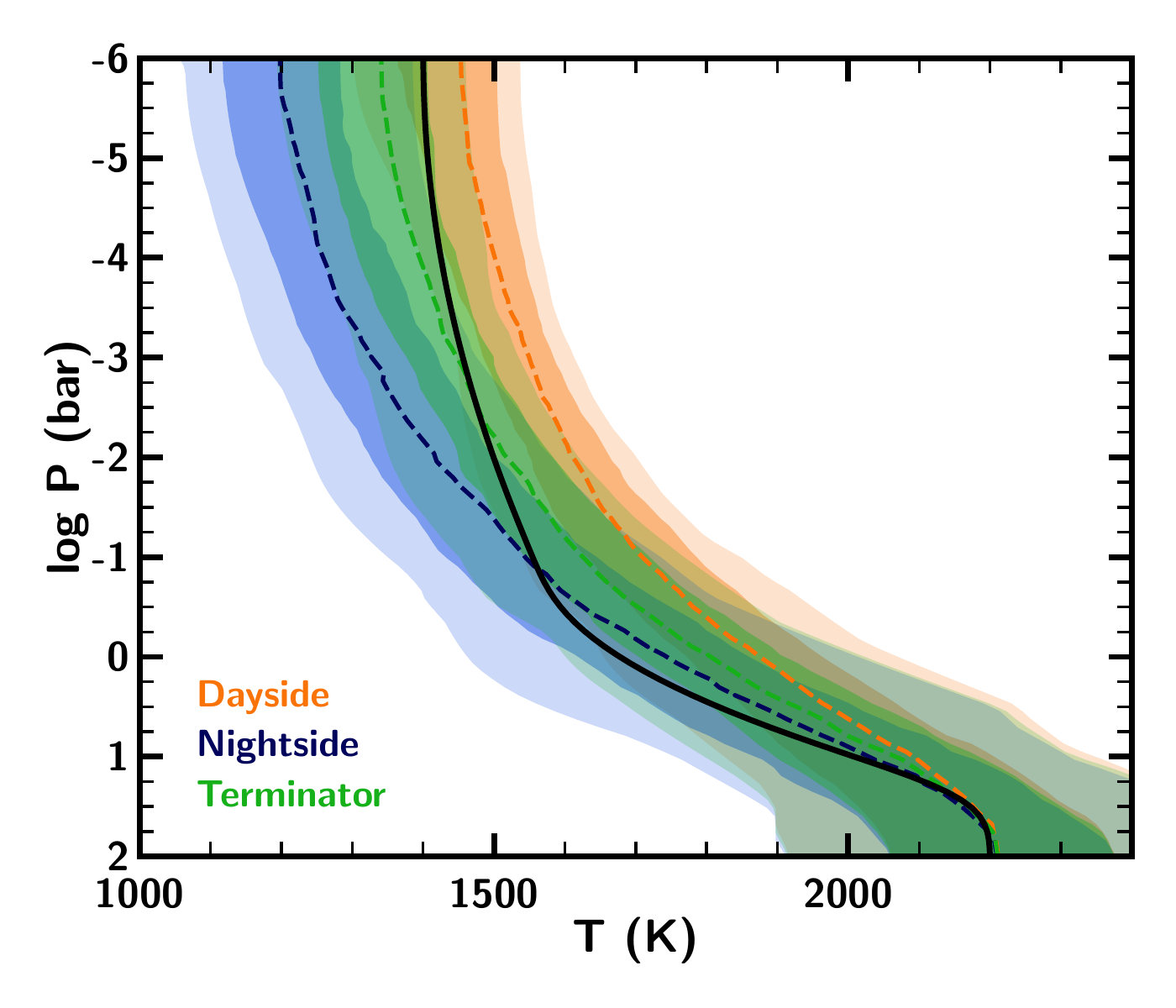}
    \caption{Input and retrieved multidimensional parametric temperature profiles of a hot Jupiter with no day-night temperature contrast. The solid line shows the input temperature profile used to generate the synthetic data. The dashed lines correspond to the median retrieved dayside, nightside and terminator temperature profiles respectively, with shaded regions corresponding to 1$\sigma$ and 2$\sigma$ confidence intervals. The retrieved temperature profiles overlap significantly with each other and are broadly consistent with the input profile, suggesting that a multidimensional retrieval approach is not required in this case.}
    \label{fig:3d_par_ret_nc}
\end{figure}

\begin{figure*}
\centering
\includegraphics[width=0.99\linewidth,trim={0 0 0 0},clip]{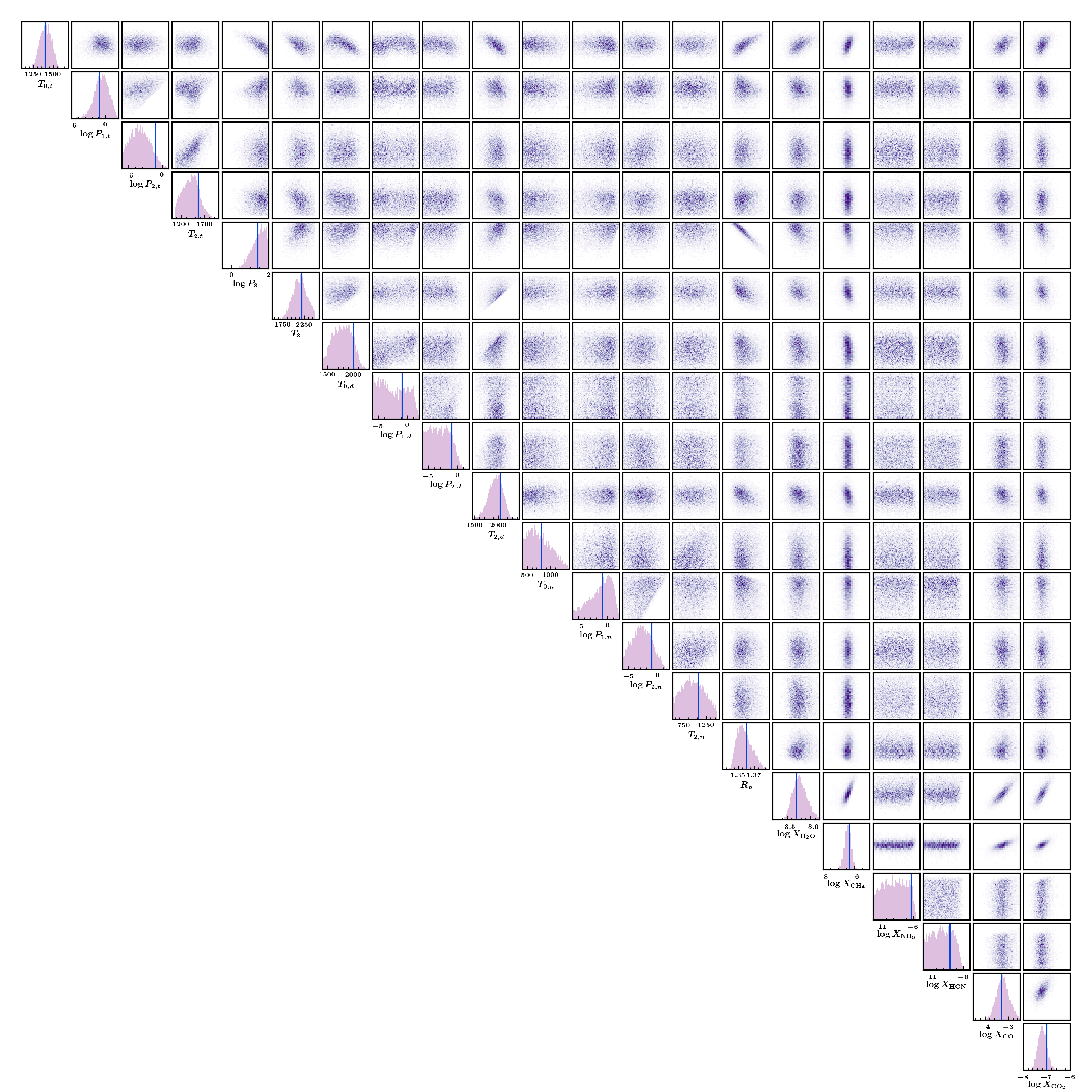}

	\def\arraystretch{1.0}
	\setlength{\tabcolsep}{5pt}
	\setlength{\arrayrulewidth}{1.3pt}
	\vspace{-10.5cm}\hspace{-11.5cm}\begin{tabular}{ccc}
		\hline
		Parameter & Input & Retrieved \\
		\hline
		$T_{0,t}$ / K & 1400 & 1406.16$^{+75.94}_{-74.96}$ \\
		$\log$ ($P_{1,t}$/bar) & -0.9 & -0.53$^{+1.07}_{-1.15}$ \\
		$\log$ ($P_{2,t}$/bar) & -1.0 & -3.37$^{+1.61}_{-1.50}$ \\
		$T_{2,t}$ / K & 1560 & 1405.43$^{+174.14}_{-191.34}$ \\
		$\log$ ($P_{3}$/bar) & 1.4 & 1.51$^{+0.32}_{-0.46}$ \\
		$T_{3}$ / K & 2200 & 2149.36$^{+170.29}_{-167.32}$ \\
		$T_{0,d}$ / K & 2000 & 1789.07$^{+191.59}_{-201.78}$ \\
		$\log$ ($P_{1,d}$/bar) & -0.9 & -2.62$^{+2.90}_{-2.37}$ \\
		$\log$ ($P_{2,d}$/bar) & -1.0 & -3.08$^{+1.95}_{-1.92}$ \\
		$T_{2,d}$ / K & 2040 & 1953.74$^{+140.69}_{-163.64}$ \\
		$T_{0,n}$ / K & 800 & 731.31$^{+302.82}_{-218.67}$ \\
		$\log$ ($P_{1,n}$/bar) & -0.9 & -0.98$^{+1.60}_{-2.58}$ \\
		$\log$ ($P_{2,n}$/bar) & -1.0 & -2.75$^{+1.79}_{-1.80}$ \\
		$T_{2,n}$ / K & 1080 & 950.89$^{+286.15}_{-267.43}$ \\
		$R_p$ / $R_J$ & 1.36 & 1.3566$^{+0.0113}_{-0.0082}$ \\
		$\log X_{\rm H_2O}$ & -3.3 & -3.23$^{+0.17}_{-0.14}$ \\
		$\log X_{\rm CH_4}$ & -6.3 & -6.41$^{+0.22}_{-0.23}$ \\
		$\log X_{\rm NH_3}$ & -6.3 & -8.80$^{+1.96}_{-2.01}$ \\
		$\log X_{\rm HCN}$ & -8.0 & -9.23$^{+1.74}_{-1.77}$ \\
		$\log X_{\rm CO}$ & -3.3 & -3.24$^{+0.29}_{-0.26}$ \\
		$\log X_{\rm CO_2}$ & -7.0 & -7.19$^{+0.20}_{-0.18}$ \\
		\hline
	\end{tabular}

    \vspace{0.3cm}\caption{Marginalised posterior probability distributions for the retrieval of a synthetic JWST transmission spectrum of the hot Jupiter HAT-P-1b. Blue vertical lines indicate true parameter values. \textit{Inset:} Input and retrieved parameter values with 1$\sigma$ uncertainties.}
    \label{fig:full_post}
\end{figure*}

\begin{figure*}
\includegraphics[width=\linewidth]{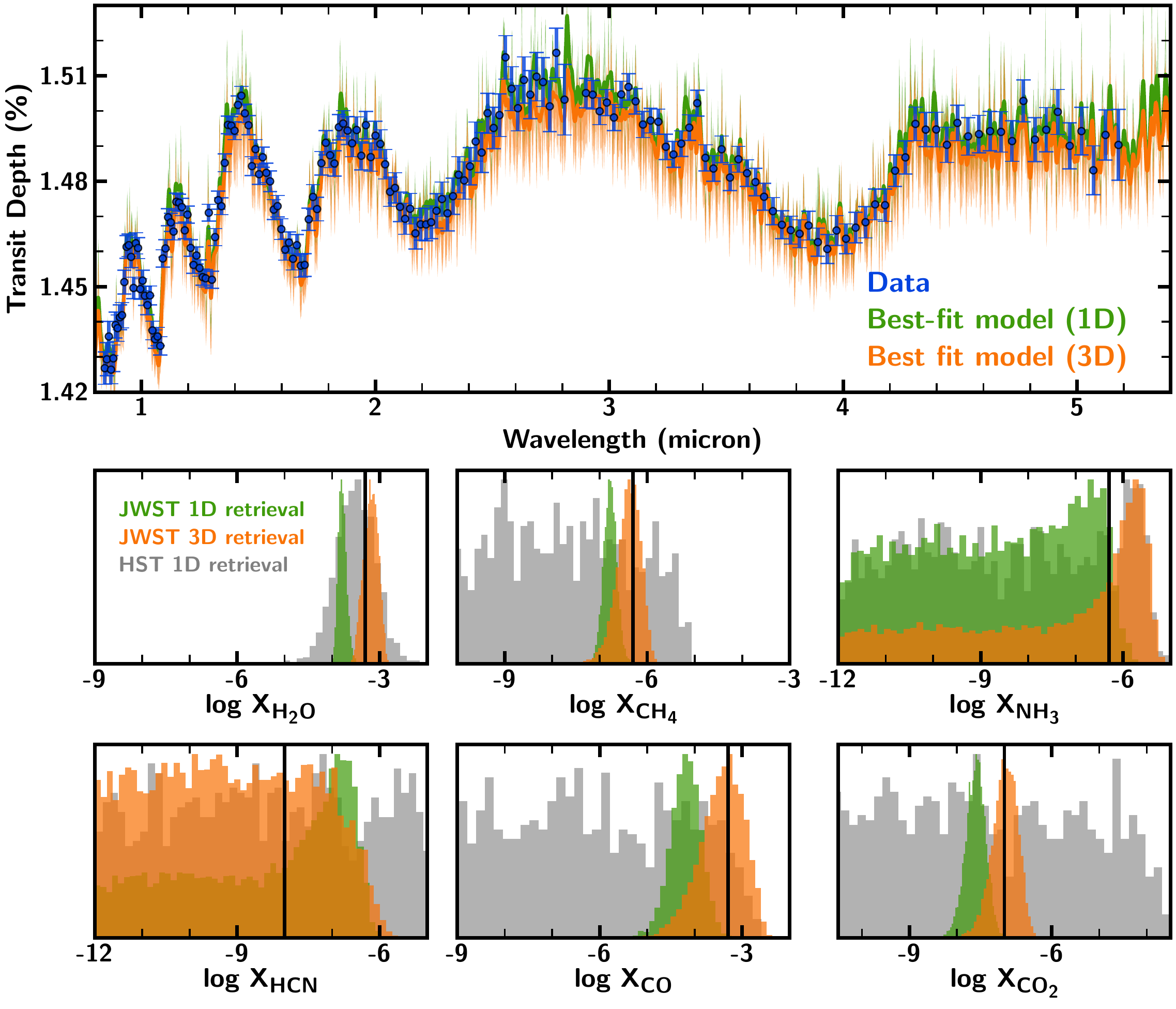}
    \caption{Atmospheric retrieval of a synthetic hot Jupiter transmission spectrum generated using a 3D temperature structure. 
    \textit{Top}: Observations (blue) and retrieved model spectra (green and orange) for the two different model considerations. Shaded regions represent 1$\sigma$ and 2$\sigma$ confidence intervals.
    \textit{Bottom}: Posterior distributions for the retrieved volume mixing ratios of each molecular species in the model. Input (equilibrium solar) values are shown by solid black lines. The model which does not include a day-night temperature contrast finds abundances that are not consistent with the input values. The model that does include a day-night temperature contrast is capable of inferring accurate abundances. The grey posterior distributions represent a retrieval of the HST WFC3 spectrum of the same model planet for comparison, using a 1D forward model.}
    \label{fig:retrieval_results}
\end{figure*}

\begin{figure*}
\includegraphics[width=\linewidth]{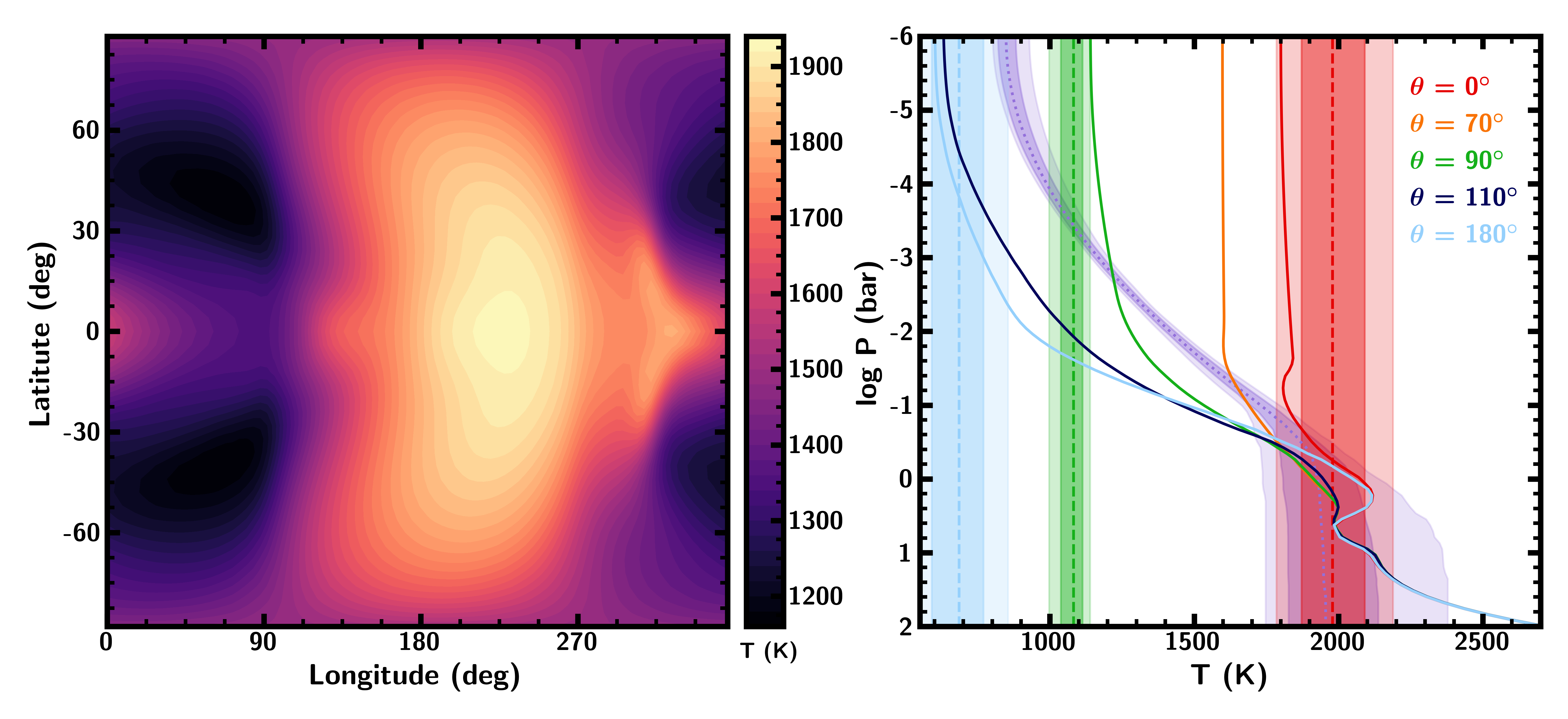}
    \caption{\textit{Left}: Temperature structure at a pressure of 0.1 bar of the forward model used to generate the synthetic spectrum. \textit{Right}: Azimuthally-averaged temperature profiles from the forward model (solid lines), along with retrieved temperature profiles from the two retrievals. The retrieved 1D $P$--$T$ profile is shown as a purple dotted line, with shaded 1$\sigma$ and 2$\sigma$ confidence intervals. The retrieved terminator (green), dayside (red) and nightside (light blue) temperatures from the retrieval incorporating day-night temperature variations are shown as dashed lines, also with shaded 1$\sigma$ and 2$\sigma$ confidence intervals.}
    \label{fig:gcm_pt_plot}
\end{figure*}

\subsection{Effect of chemical inhomogeneity on transmission spectra} \label{subsec:chem}

Since \textsc{Aura-3D} is capable of modelling atmospheres with inhomoegeneous chemistry, we explore the possibility for chemical variability between the day- and nightsides of a planet to affect resulting transmission spectra. Past studies have demonstrated that thermal dissociaton of molecules such as H$_2$O on the daysides of ultra-hot Jupiters can substantially affect transmission spectra \citep{Lothringer2018,Parmentier2018,Pluriel2020,Pluriel2022}. In this section we consider a different chemical transition, namely the transition from a CH$_4$-dominated to CO-dominated atmosphere. In chemical equilibrium, the abundances of key atmospheric species in a H$_2$-rich atmosphere can vary strongly with temperature. At temperatures below $\sim$1200~K, CH$_4$ is expected to be a highly abundant molecule, whereas at higher temperatures CO is expected to dominate \citep{Lodders2002,Moses2011,Madhu2012}. If a planet's day- and nightsides are above and below this temperature respectively, this could lead to substantially different chemical abundances on either side of the planet.

In order to investigate this effect, we generate models of the hot Neptune HAT-P-11b \citep{Bakos2010}. Given this planet's equilibrium temperature of 878$\,$K, it is possible that the day- and nightsides of the planet are above and below the transition temperature between CH$_4$-dominated and CO-dominated chemistry, making the planet a useful test case to examine the effects of chemical inhomogeneity on transmission spectra. We consider two different sets of equilibrium chemical abundances: a CO-dominated regime, which we may expect to see on the dayside of the planet, and a CH$_4$-dominated regime which may be expected on the nightside. The abundances for each regime are shown in Table \ref{tab:abundances}. We choose a model $P$--$T$ profile with a similar structure to the profiles used in Section \ref{subsec:mgrid}. We take $T_t=878\,$K, the equilibrium temperature of the planet, with $\Delta T=500\,$K. The values of $\log P_1$, $\log P_2$ and $\log P_3$ are again fixed to -0.9, -1.0 and 1.4 respectively. For the chemistry we consider four different scenarios: (1) dayside abundances across the whole terminator region, (2) nightside abundances across the whole terminator, (3) inhomogeneous chemistry with dayside abundances for $\theta < 90^{\circ}$ and nightside abundances for $\theta > 90^{\circ}$, and (4) averaged abundances across the whole terminator (the mean of the dayside and nightside abundances). We generate models from 0.5-15$\,\mu$m, which covers the wavelength ranges of several JWST instruments, including NIRSpec, NIRISS NIRCam, and MIRI LRS, and also covers many of the prominent spectral features of the molecules considered \citep{Madhu2019}.

The resulting spectra are shown in Figure \ref{fig:daynight_spectra}. We see that the model with nightside chemistry across the entire terminator has a higher transit depth overall than the model with dayside chemistry across the entire terminator. This is a result of the high abundances of H$_2$O, CH$_4$ and NH$_3$ expected in chemical equilibrium at low temperatures, each of which has prominent absorption features at the wavelengths considered here. The model with averaged chemistry also differs substantially from the model with chemical abundances varying between the day- and nightsides of the planet across the wavelength range considered. The model with averaged chemistry has larger transit depths, since this model has large amounts of H$_2$O, CH$_4$ and NH$_3$ on the dayside as well as the nightside, and the dayside has a larger scale height since it is at a higher temperature. In the model with variable chemistry, these molecules (particularly CH$_4$ and NH$_3$) are mostly present on the cooler side of the planet, leading to smaller features overall.

Whether such chemical inhomogeneities can be detected with JWST will need to be determined in future studies. For the remainder of this work we will focus on the possibility of constraining temperature inhomogeneities using JWST-quality observations.

\begin{table}
	\centering
	\caption{Chemical abundances used for the day- and nightsides of our HAT-P-11b models. The values shown are the logarithm of the volume mixing ratios.}
	\hfill \\
	\label{tab:abundances}
	\def\arraystretch{1.5}
	\setlength{\arrayrulewidth}{1.3pt}
	\begin{tabular}{ccc}
		\hline
		Species & Dayside abundance & Nightside abundance \\
		\hline
		H$_2$O & -3.3 & -3.0  \\
		CH$_4$ & -6.3 & -3.3 \\
		NH$_3$ & -6.3 & -3.6  \\
		HCN & -8.0 & -20.0 \\
		CO & -3.3 & -16.0 \\
		CO$_2$ & -7.0 & -17.0 \\
		\hline
	\end{tabular}
\end{table}

\subsection{Retrieval of a multidimensional temperature structure} \label{subsec:parametric_ret}

Here we demonstrate the capability of our new retrieval framework to recover a multidimensional temperature structure. We create a parametric temperature profile using the prescription described in Section \ref{subsubsec:3dpt}. We use this temperature structure to produce a synthetic transmission spectrum, and carry out a retrieval of this spectrum using the same parameterisation. Our goal is to determine the extent to which the retrieval can constrain the input multi-dimensional temperature structure using JWST-quality data.

The input parameters used to generate the temperature structure are shown in the table embedded in Figure \ref{fig:full_post}. The model interpolates between temperature profiles defined for the dayside ($\theta \leq 70^{\circ}$), nightside ($\theta \geq 110^{\circ}$) and the middle of the terminator ($\theta = 90^{\circ}$). We consider a cloud-free, solar composition atmosphere with uniform abundances of H$_2$O, CH$_4$, NH$_3$, HCN, CO and CO$_2$. Input chemical abundances are also shown in the table embedded in Figure \ref{fig:full_post}.

To generate our synthetic data set we assume planetary and stellar properties that are similar to the hot Jupiter HAT-P-1b \citep{Bakos2010}. We choose this planet as an intermediate case that is representative of the majority of hot Jupiters that may be observed with JWST. HAT-P-1b has similar bulk properties to HD~209458b, and we find day-night temperature gradients to produce similar differences in the transmission spectrum to those shown in Figure \ref{fig:spectrum_comparison}. The forward model is first computed at a moderately high resolution ($R = 3000, 0.5-5.5 \mu$m). Synthetic observations are then generated using PANDEXO \citep{Batalha2017}, assuming a single transit observed with NIRISS SOSS and another single transit observed with NIRSpec G395H. This provides a broad wavelength coverage with JWST that contains multiple spectral features of the chemical species \citep[][]{Madhu2019} included in the forward model. The Nested Sampling retrieval uses 2000 live points.

The input data and retrieved spectrum are shown in Figure \ref{fig:3d_par_ret_spec}. We find excellent agreement between the input and retrieved spectra across the full wavelength range considered. Marginalised posterior distributions for each model parameter, as well as the retrieved median values with associated 1$\sigma$ uncertainties, are shown in Figure \ref{fig:full_post}. The estimates of most free parameters are consistent with input values to within 1$\sigma$ uncertainty, and all retrieved parameters are consistent with input values to within 2$\sigma$.

Our input and retrieved temperature profiles are shown in Figure \ref{fig:3d_par_ret}. Each of the dayside, nightside and terminator input temperature profiles almost entirely remain within the retrieved $1\sigma$ confidence intervals of the retrieved profiles, and lie completely within the $2\sigma$ confidence intervals. The uncertainty estimates for the temperature profiles in the model range from $\sim$100--300$\,$K. We also compare 2D representations of the input and median retrieved temperature structures in Figure \ref{fig:polar_pt_ret}. We see that the temperature structures are very similar, highlighting the retrieval's ability to recover a detailed multidimensional temperature profile using high-quality data.

The retrieved abundances of all chemical species are shown in Figure \ref{fig:full_post}. The abundances of H$_2$O, CH$_4$, CO and CO$_2$ are all constrained to within 0.5 dex and are consistent with input values to 1$\sigma$. We obtain upper limits on the abundances of NH$_3$ and HCN which are also consistent with input values, noting that these species have low abundances in the input model ($\log X_{\rm NH_3}=-6.3, \log X_{\rm HCN}=-8.0$).

This result demonstrates that using JWST observations of a nominal hot Jupiter, it is possible to obtain accurate constraints on a multidimensional temperature profile across the terminator region with reasonable precision. Furthermore, these observations can be used to obtain very precise abundance estimates for a number of chemical species.

As an additional validation test for our retrieval pipeline, we also conduct a retrieval of a synthetic spectrum for a planet with no day-night temperature gradient (i.e. a 1D $P$--$T$ profile) using our 3D framework. The retrieved dayside, nightside and terminator temperature profiles are shown in Figure \ref{fig:3d_par_ret_nc}. We find that there is considerable overlap between each of the retrieved profiles, as would be expected for a planet with no temperature contrast. In this case, the lack of strong distinction between the three profiles suggests that this spectrum is better suited to a 1D retrieval (see Section \ref{subsec:1d3d}).

\subsection{Retrieval of a synthetic spectrum from a GCM} \label{subsec:gcm_ret}

In this section we present retrievals of a simulated JWST transmission spectrum of a hot Jupiter generated with a 3D temperature structure. We conduct retrievals using two different modelling paradigms: one using a 1D parametric $P$-$T$ profile, and one using a temperature profile that is allowed to vary between the day- and nightsides of the planet.

The temperature structure of our input forward model is adapted from the open-source General Circulation Model THOR \citep{Mendonca2016,Deitrick2020}. We use THOR to reproduce the "Deep Hot Jupiter" scenario described in \citet{Deitrick2020}. In order to create a temperature structure suitable for our forward model, we subsequently extrapolate the output temperature profiles from the GCM to lower pressures. In order to do this, we fit a parabolic $P$-$T$ curve to each temperature profile in the longitude/latitude grid of outputs. The resulting 3D $P$-$T$ structure is shown in Figure \ref{fig:gcm_pt_plot}.

Having adapted the GCM temperature structure to be appropriate for our retrieval case study, we now use it to create a model JWST transmission spectrum. Our forward model in this instance is cloud-free and assumes chemical abundances of a 1$\times$solar composition H$_2$-dominated atmosphere \citep[see][]{Gandhi2017}, including H$_2$O, CH$_4$, NH$_3$, HCN, CO and CO$_2$. We assume the same planetary and stellar properties for the planet HAT-P-1b as in Section \ref{subsec:parametric_ret}.

We carry out two retrievals of our simulated JWST spectrum. The first such retrieval uses the 1D parametric $P$-$T$ prescription as described in \citet{Madhu2009}, while the second uses a temperature profile that varies across the terminator, but not with height (see Section \ref{subsubsec:thetapt}). We fix $\beta$ to 40$^{\circ}$, which is representative of the extent of the terminator region which affects a transmission spectrum \citep{Fortney2010}. We do not allow cases in which the dayside temperature is colder than the terminator temperature, or where the nightside temperature is hotter than the terminator.

A comparison of the two retrieval results is shown in Figure \ref{fig:retrieval_results}. We find that the retrieval assuming a 1D parametric $P$--$T$ profile leads to some inaccuracies in the retrieved abundances using JWST-quality data. None of the retrieved abundances found using the 1D model are consistent with the input values to within 1$\sigma$. The only abundance retrieved accurately to $\leq$2$\sigma$ is that of HCN, while the input abundance of NH$_3$ is within 3$\sigma$ of the retrieved posterior. Note that these abundances are also the least well-constrained of the species considered. All other abundances fall more than 3$\sigma$ away from the retrieved posteriors.

This finding is similar to that of \citet{Caldas2019}, who found that when applying a 1D retrieval framework to a high-quality spectrum generated with a 3D input model, the true abundance would often be outside the error bars of the retrieval. They also find that the magnitude and direction of the error depends on the specific case; in this particular example, the retrieved abundances of most species are lower than the true values.

In contrast, for the retrieval in which the temperature is allowed to vary across the terminator, each of the retrieved chemical abundances is consistent within 1$\sigma$ of the true values. Furthermore, the uncertainties on the parameter estimates in this case are generally larger than in the 1D retrieval. This suggests that the additional flexibility introduced by incorporation of a day-night temperature gradient allows the model to explore a wider range of models that can explain the data at hand. This demonstrates that our prescription is able to mitigate against biases that can arise when only a 1D temperature profile is allowed in a retrieval of JWST-quality data.

We also carry out a retrieval of a simulated HST WFC3 spectrum of the same planet for comparison, using a 1D atmospheric model. In this case, the only well-constrained abundance is that of H$_2$O, with a precision of 0.8 dex. The retrieved abundance of H$_2$O is consistent with the input value, albeit with a larger uncertainty. This suggests that the constraints obtained from retrievals of HST data are not sufficiently precise to necessitate a multidimensional retrieval algorithm.

The retrieved temperature profiles acquired from each retrieval are shown on the right-hand panel of Figure \ref{fig:gcm_pt_plot}. The retrieved 1D temperature profile exhibits a strong temperature gradient with pressure that is not found at the terminator of the input model. For the model including a day-night temperature gradient, the retrieved temperatures are broadly consistent with the dayside, nightside and terminator temperatures in the upper atmosphere. We note that both of these prescriptions are heavily simplified in comparison to a full 3D temperature structure. Therefore, it is difficult to gain complete information about the temperature structure of the planet's atmosphere from these retrievals alone, beyond the fact that a strong day-night temperature gradient appears to exist. However, this example demonstrates importance of having the capability to include a day-night temperature gradient in retrievals in order to ensure that we can obtain accurate abundance estimates when working with high-quality spectra of hot Jupiters.

\begin{table}
	\centering
	\caption{Description of priors for retrievals of simulated JWST transmission spectra of HAT-P-1b generated using a GCM. The temperature profile for the 1D retrieval uses the first six parameters ($T_0, \alpha_1, \alpha_2, P_1, P_2, P_3$) while the retrieval with a day-night temperature gradient uses the parameters $T_0, T_{0,\rm{d}}, T_{0,\rm{n}}$.}
	\hfill \\
	\label{tab:priors_jwst}
	\def\arraystretch{1.5}
	\setlength{\arrayrulewidth}{1.3pt}
	\begin{tabular}{cccc}
		\hline
		Parameter & Lower Bound & Upper Bound & Prior \\
		\hline
		$T_{\textnormal{0}}$ (K) & $600$ & $2200$ & uniform \\
		$\alpha_1$ & 0.02 & 1.0 & uniform \\
		$\alpha_2$ & 0.02 & 1.0 & uniform \\
		$P_{\textnormal{1}}$ (bar) & $10^{-6}$ & $10^{2}$ & log-uniform \\
		$P_{\textnormal{2}}$ (bar) & $10^{-6}$ & $10^{2}$ & log-uniform \\
		$P_{\textnormal{3}}$ (bar) & $10^{-2}$ & $10^{2}$ & log-uniform \\
		$T_{\textnormal{0,d}}$ (K) & $1200$ & $2600$ & uniform \\
		$T_{\textnormal{0,n}}$ (K) & $400$ & $1800$ & uniform \\
		$X_{i}$ & $10^{-12}$ & $10^{-2}$ & log-uniform\\
		\hline
	\end{tabular}
\end{table}

\begin{table}
	\centering
	\caption{Input and retrieved abundances for the simulated JWST transmission spectrum of HAT-P-1b. The abundances are shown as the logarithm of the volume mixing ratios.}
	\hfill \\
	\label{tab:results_jwst}
	\def\arraystretch{1.5}
	\setlength{\arrayrulewidth}{1.3pt}
	\begin{tabular}{cccc}
		\hline
		Species & True value & No $T$ gradient & $T$ gradient \\
		\hline
		H$_2$O & -3.3 & -3.78$^{+0.09}_{-0.08}$ & -3.18$^{+0.15}_{-0.16}$  \\
		CH$_4$ & -6.3 & -6.78$^{+0.13}_{-0.14}$ & -6.38$^{+0.23}_{-0.28}$ \\
		NH$_3$ & -6.3 & -8.72$^{+1.89}_{-2.16}$ & -6.95$^{+1.24}_{-3.33}$ \\
		HCN & -8.0 & -8.10$^{+1.34}_{-2.57}$ & -9.21$^{+1.93}_{-1.86}$ \\
		CO & -3.3 & -4.18$^{+0.30}_{-0.32}$ & -3.39$^{+0.43}_{-0.50}$ \\
		CO$_2$ & -7.0 & -7.60$^{+0.20}_{-0.22}$ & -6.98$^{+0.25}_{-0.28}$ \\
		\hline
	\end{tabular}
\end{table}


\section{Summary and Discussion} \label{section:discussion}

We introduce \textsc{Aura-3D}, a 3D modeling and retrieval framework for transmission spectra of exoplanetary atmospheres. The high data quality expected from upcoming JWST observations of exoplanet transmission spectra motivated the development of this new framework. Our new forward model enables calculation of transmission spectra with any 3D temperature structure, including GCM outputs, as well as computation of model spectra using parametric temperature profiles that include day-night inhomogeneities. Here we summarize the key functionalities of \textsc{Aura-3D} retrieval framework. 
\begin{itemize}
    \item The framework includes a 3D parametric temperature profile for use in atmospheric retrieval that can fit a wide range of GCM temperature structures. This parameterisation enables the computation of transmission spectrum models with highly flexible, physically realistic temperature structures incorporating thermal inhomogeneities. 
    \item The radiative transfer in 3D geometry is computationally efficient for retrievals, with computation time for a single model spectrum in 3D geometry $\lesssim$1s. We have demonstrated retrievals using a 3D geometry requiring over 10$^6$ model evaluations. This is easily scalable to higher order depending on the requirement.
    \item When applied to synthetic JWST data of hot Jupiters, \textsc{Aura-3D} is able to accurately constrain separate dayside, nightside and terminator temperature profiles probed in transmission geometry, yielding meaningful information on day-night temperature contrasts.
    \item \textsc{Aura-3D} can be used to accurately retrieve chemical abundances from transmission spectra whose temperature structures are calculated with 3D GCMs, even with JWST-quality data of hot Jupiters where 1D retrievals may lead to biased estimates.
\end{itemize}
We use \textsc{Aura-3D} to investigate constraints that can be placed on hot giant exoplanets using JWST-quality spectra, finding the following results:
\begin{itemize}
    \item For hot Jupiters with photospheric terminator temperatures $\gtrsim$1100 K and temperature contrasts $\gtrsim$500 K, the day-night temperature gradient can cause differences in transmission spectra that could be detectable with JWST. 
    \item We demonstrate that it is possible to constrain a multidimensional temperature profile across the day-night terminator of a nominal hot Jupiter to a precision of $\sim$100-300$\,$K using nominal JWST-quality data.
    \item We demonstrate that for JWST spectra of some hot Jupiters, a 1D retrieval can lead to biased abundance estimates, in agreement with previous studies. However, we find that in the case presented in this work, the introduction of a multidimensional temperature profile can overcome these biases and retrieve accurate abundances.
    \item For atmospheres where the terminator may transition between CO/CH$_4$ dominated composition, models with inhomogeneous composition may cause an observable difference to transmission spectra compared to those with globally-averaged compositions.
\end{itemize}

\subsection{Applicability of 1D vs 3D retrievals} \label{subsec:1d3d}

The developments presented in this work allow for retrievals with a much more complex forward model than has previously been possible. Our new framework will be extremely useful for analysing JWST-quality spectra of hot Jupiters with inhomogeneous terminators. While it is tempting to apply the most complex forward model available to all retrievals, care needs to be taken in ascertaining whether a 3D framework is necessary to explain a given observed spectrum. Furthermore, simply starting from the most complex possible model may not be computationally feasible. Although we have shown that retrievals with a 3D geometry can be carried out with high computational efficiency, the combination of 3D modelling alongside inhomogeneous clouds/hazes, stellar heterogeneity, large numbers of chemical species and other possible features such as inhomogeneous chemistry would lead to incredibly computationally expensive retrievals with an extremely high number of free parameters. It is therefore prudent to asses a priori what degree of complexity is required for a given spectrum, finding an optimal parameter set for analysing the data at hand.

While we have demonstrated that thermal inhomogeneities will lead to substantial deviations between 1D and 3D models for hot giant planets with large day-night temperature contrasts, we also find that for cooler planets the effect of thermal inhomogeneities is smaller and, therefore, may not be detectable even with JWST-quality data. When analysing future observations, several properties of the planet should be taken into account when considering the best $P$--$T$ profile prescription, including the planetary radius, gravity and equilibrium temperature, all of which affect the amplitude of spectral features and therefore the degree to which effects such as day-night temperature gradients will be detectable. The extent of the terminator region being probed in transmission is also important to consider. Hotter planets tend to have larger opening angles, allowing for a greater variation in atmospheric conditions across the region probed in transmission \citep{Wardenier2022}. When considering retrievals of real observed data it will be important to determine whether the opening angle is sufficiently large to justify the inclusion of a 3D model.

As well as the planet being targeted, it is important to consider the quality of the observed data when determining the appropriate retrieval paradigm. This study has largely focused on high-quality JWST observations with a broad infrared wavelength coverage. We have found that deviations between 1D and 3D models are less significant when considering HST-quality data. Furthermore, other studies have demonstrated that the assumption of a 1D temperature profile will generally not lead to biases when analysing HST transmission spectra \citep{Welbanks2022} of most hot Jupiters. For future observations, the optimal retrieval paradigm will need to be chosen taking into account the wavelength range being probed as well as the resolution and precision of the spectrum.

\subsection{Complementary observations using emission spectra and phase curves}

In an ideal scenario, in order to obtain complete information about the structure and properties of an exoplanet's atmosphere, we would acquire its transmission and emission spectra as well as its phase curve. This combination of observations would allow for detailed characterisation of a planet's terminator region through transmission spectroscopy and of its dayside through emission spectroscopy, as well as providing insight into spatially-resolved properties through phase curve analysis. However, undertaking such an extensive observing campaign for a single planet would require very large amounts of telescope time, and if the planet does not exhibit strong spatial inhomogeneities, then time-consuming phase curve observations are unlikely to yield additional insights into the nature of its atmosphere.

The results from this work indicate that it should be possible to detect thermal inhomogeneities in certain exoplanet atmospheres using only transmission spectroscopy. We therefore suggest that an optimal observing strategy would be to first observe a planet's transmission spectrum to look for hints of day-night variability, possibly complementing this observation with an emission spectrum to constrain dayside properties and break certain degeneracies (see Section \ref{subsec:future}). Using these observations it will be possible to ascertain the degree to which spatial variations in temperature and chemistry permeate the planet's atmospheric structure, thus determining whether observations of the full phase-curve will lead to further meaningful results.

\subsection{Future developments} \label{subsec:future}

The retrieval framework presented here enables 3D atmospheric characterisation using transmission spectra, a crucial step towards robust analysis of JWST data. However, there are additional atmospheric properties that will be important to consider when interpreting upcoming observations. For example, while inhomogeneous cloud/haze coverage has been considered in past works \citep[e.g.,][]{Line2016,MacDonald2017,Welbanks2021}, the combined effect of patchy clouds/hazes and inhomogeneous temperature profiles is yet to be fully explored in a retrieval context.

The parametric $P$--$T$ profile incorporated into \textsc{Aura-3D} has been sufficiently complex to analyse the synthetic JWST spectra considered in this study. \textsc{Aura-3D} has the capability to model thermal variations with both $\theta$ and $\phi$, however only variations with $\theta$ have been explored to carry out the retrievals presented in this work. However, for certain targets it is possible that an even higher signal-to-noise will be achieved, either due to observations of atmospheres with a very large scale height or via observing multiple transits of a given target. In these cases it is possible that more complexity, including a $\phi$-dependent temperature profile, will be required to accurately retrieve certain parameters. Our current framework has a built-in $\phi$-dependence which has already been used to model patchy clouds \citep{Pinhas2019}, and has the functionality to include a parametric temperature profile with variations in both $\phi$ and $\theta$.

It is also important to assess the impact of inhomogeneous chemical abundances on retrievals of transmission spectra. This topic has been explored for ultra-hot Jupiters \citep{Pluriel2022} but may also be relevant for cooler planets, as discussed in this work. While \textsc{Aura-3D} is capable of modelling planets with different day- and nightside chemical compositions, further work is needed to ascertain the feasibility of retrieving inhomogenous chemical abundances directly from a transmission spectrum. A retrieval allowing for independent dayside and nightside chemical abundances for each species would be subject to very strong degeneracies, since low dayside abundances could be compensated for by high nightside abundances and vice-versa. It would therefore be important to limit the allowed parameter space to physically realistic dayside and nightside abundances motivated by detailed self-consistent models. As discussed above, combining transmission spectra with observations of other regions of a planet's atmosphere, such as emission spectra, could also be useful in breaking this degeneracy.

Ultimately, we should strive to incorporate the most sophisticated forward models possible into retrieval algorithms, so that they can be invoked if necessary to explain observations. Any additional increases in model complexity must be accompanied by concurrent efforts to make these models efficient for retrievals, as we have done in the present work. Efforts are underway to improve the computational efficiency of forward models and retrievals through means such as GPU implementations \citep[e.g.,][]{Malik2017} and Machine Learning \citep[e.g.,][]{MarquezNeila2018,Nixon2020}. These avenues for improvement may enable further steps towards reducing other simplifying assumptions that still remain in place in current retrieval frameworks.

The developments presented in this work represent an important step forwards in our ability to identify and constrain multidimensional effects in exoplanet atmospheres. This opens the field to comprehensive atmospheric characterisation of exoplanets with large spectral features that will be accessible with transmission spectroscopy in the JWST era. Our retrieval framework will be an important tool as we work to uncover exciting new insights into the nature of exoplanet atmospheres using upcoming observations of unprecedented detail.

\section*{Acknowledgements}

The authors thank the anonymous referee for their helpful comments, which improved the quality of the manuscript. MN and NM acknowledge support from the Cambridge Centre for Doctoral Training in Data Intensive Science, which is funded by the Science and Technology Facilities Council (STFC), UK. High-Performance Computing resources required to carry out this work were provided by the Cambridge Service for Data Driven Discovery (CSD3), operated by the University of Cambridge Research Computing Service (www.csd3.cam.ac.uk), provided by Dell EMC and Intel using Tier-2 funding from the Engineering and Physical Sciences Research Council (capital grant EP/P020259/1) and DiRAC funding from STFC.  MN thanks Stephen Thorp for helpful discussion regarding Figure \ref{fig:polar_pt}. This research has made use of the NASA Astrophysics Data System and the Python packages \textsc{numpy} \citep{Harris2020}, \textsc{scipy} \citep{Virtanen2020}, \textsc{matplotlib} \citep{Hunter2007} and \textsc{starry} \citep{Luger2019}.



\bibliographystyle{aasjournal}
\bibliography{main}

\label{lastpage}
\end{document}